\documentclass[iop]{emulateapj}

\usepackage{graphicx,amsmath}
\usepackage{times}

\DeclareMathOperator\erf{erf}

\shorttitle{Magnetic Flux Transport and the Long-Term Evolution of Solar Active Regions}
\shortauthors{Ugarte-Urra, Upton, Warren \& Hathaway}

\begin{document}


\title{Magnetic Flux Transport and the Long-Term Evolution of Solar Active Regions}
\author{Ignacio Ugarte-Urra\altaffilmark{1}, Lisa Upton\altaffilmark{2}, Harry
  P. Warren\altaffilmark{3}, and David H. Hathaway\altaffilmark{4}}

\affil{\altaffilmark{1}College of Science, George Mason University. 4400 University Drive,
  Fairfax, VA 22030, USA}
\affil{\altaffilmark{2}Independent Researcher, Broomfield, CO, 80020, USA}
\affil{\altaffilmark{3}Space Science Division, Code 7681, Naval Research Laboratory, Washington,
  DC 20375, USA}
\affil{\altaffilmark{4} NASA Ames Research Center, Mail Stop 258-5, Moffett Field, CA 94035, USA}


\begin{abstract}
 With multiple vantage points around the Sun, STEREO and SDO imaging observations provide a unique
 opportunity to view the solar surface continuously.  We use \ion{He}{2} 304\,\AA\ data from these
 observatories to isolate and track ten active regions and study their long-term evolution. We
 find that active regions typically follow a standard pattern of emergence over several days
 followed by a slower decay that is proportional in time to the peak intensity in the
 region. Since STEREO does not make direct observations of the magnetic field, we employ a
 flux-luminosity relationship to infer the total unsigned magnetic flux evolution. To investigate
 this magnetic flux decay over several rotations we use a surface flux transport model, the
 {\it Advective Flux Transport} (AFT) model, that
 simulates convective flows using a time-varying velocity field and find that the model provides
 realistic predictions when information about the active region's magnetic field strength and
 distribution at peak flux is available.  Finally, we illustrate how 304 \AA\ images can be used
 as a proxy for magnetic flux measurements when magnetic field data is not accessible.
\end{abstract}


\section{Introduction}

Solar activity is strongly modulated by variations in the Sun's surface magnetic fields. The
number of flares and coronal mass ejections, for example, varies in phase with the Sun's 11-year
sunspot cycle. Similarly, the magnitude of the Sun's radiative output also changes as the amount
of magnetic flux on the solar surface varies. Thus understanding the distribution and evolution of
surface magnetic fields is central to discovering the physical mechanisms responsible for the
heating of the solar corona and the initiation of coronal mass ejections as well as for developing
models that accurately forecast space weather.

The first numerical model of magnetic surface flux transport was developed by
\citet{DeVore_etal84} and during the past several decades there has been considerable progress in
simulating the evolution of surface magnetic fields \citep[for reviews
  see][]{Sheeley2005,Mackay2012}. These models include the effects of differential rotation,
supergranular diffusion, meridional flow, and flux emergence and cancellation. Much of this past
work has focused on very long time-scales and understanding the transport mechanisms
that give rise to the solar cycle. Of particular importance is the role of meridional flow in
carrying magnetic flux to the pole where it cancels the existing polar flux and creates the new
poloidal field that will seed the next solar cycle.

Our understanding of magnetic flux transport over shorter time scales is, however, much less
developed. Because of line-of-sight effects, magnetic observations are generally only reliable
while active regions are within a few days of central meridian. With the recent launches of the
Solar TErrestrial RElations Observatory \citep[STEREO,][]{Kaiser2008} and the Solar Dynamics
Observatory \citep[SDO,][]{Pesnell2012}, we have a unique opportunity to view the entire
solar surface continuously for long periods of time. The twin STEREO spacecraft orbit the Sun at
a distance of about 1\,AU, but drift away from the earth (in opposite directions) at a rate of
about 22${^\circ}$ per year. SDO has a geosynchronous orbit that allows for almost continuous
observations of the Sun from Earth.

In this paper we investigate three questions: [1] How does the total magnetic flux in solar active
regions evolve over a period of several solar rotations? [2] Can surface flux transport
simulations accurately model this short-term evolution? and [3] What information is necessary to
forecast the near-term evolution of solar active regions from far-side observations?

To address these questions we combine STEREO and SDO observations to make 360$^\circ$ maps of the
Sun. The absence of magnetographs on the STEREO spacecraft means that we cannot study the
evolution of the magnetic field directly with these observations. However, both STEREO and SDO
image the Sun in \ion{He}{2} 304\,\AA\ and we use this channel as a proxy for the total unsigned
magnetic flux. As is well known, the radiance in this line is well correlated with the strength of
the magnetic field. This is an example of the ``flux-luminosity'' relationship
\citep[e.g.,][]{schrijver1987}.

We isolate and track individual active regions and create continuous light curves for the active
regions of interest. We find that the total active region intensity generally follows a similar
pattern of emergence over several days followed by a decay with a lifetime proportional to the
peak flux in the region. To model this evolution we employ the surface flux transport model
recently developed by \cite{UptonHathaway14A, UptonHathaway14B}. This is the first global model to
simulate the advection of the magnetic field using a time-varying velocity field instead of an
implied diffusion rate. We find that the observed evolution is generally well reproduced by the
simulations. Finally, we illustrate how the total flux and geometrical parameters of newly emerged
active regions can be estimated from far-side images and embedded in surface flux transport
simulations to forecast their near-term evolution. This is an important first step in developing
accurate space weather forecasting tools.

\begin{figure*}[htbp!]
\centering
\includegraphics[angle=90,width=\columnwidth]{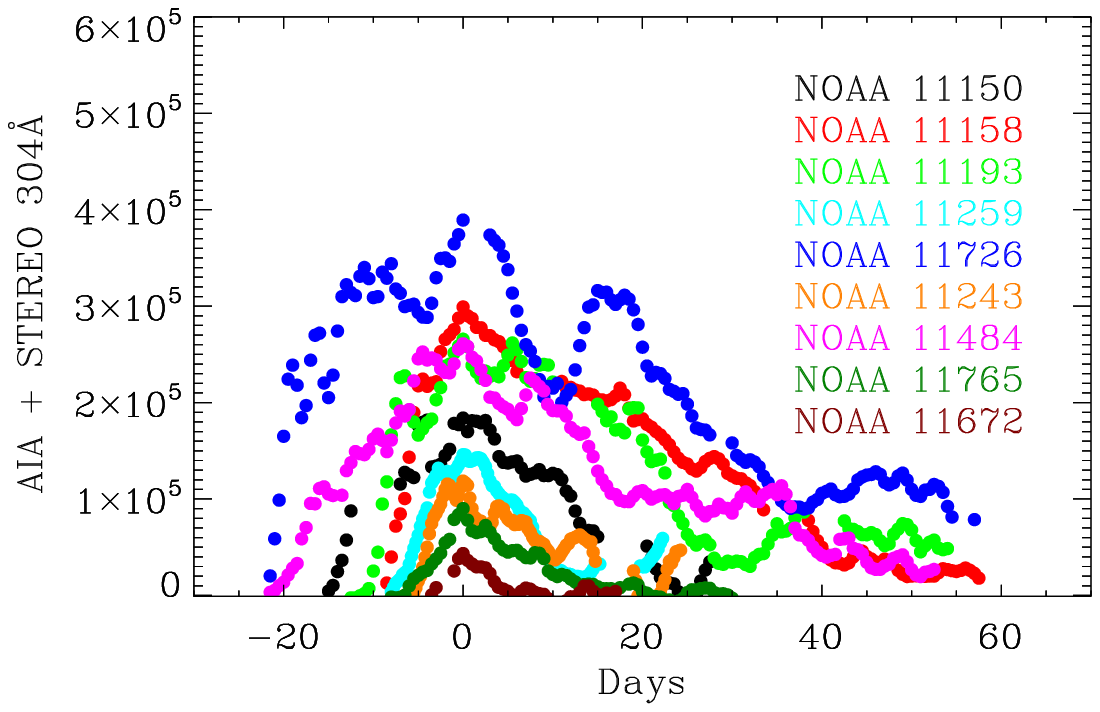}
\includegraphics[angle=90,width=\columnwidth]{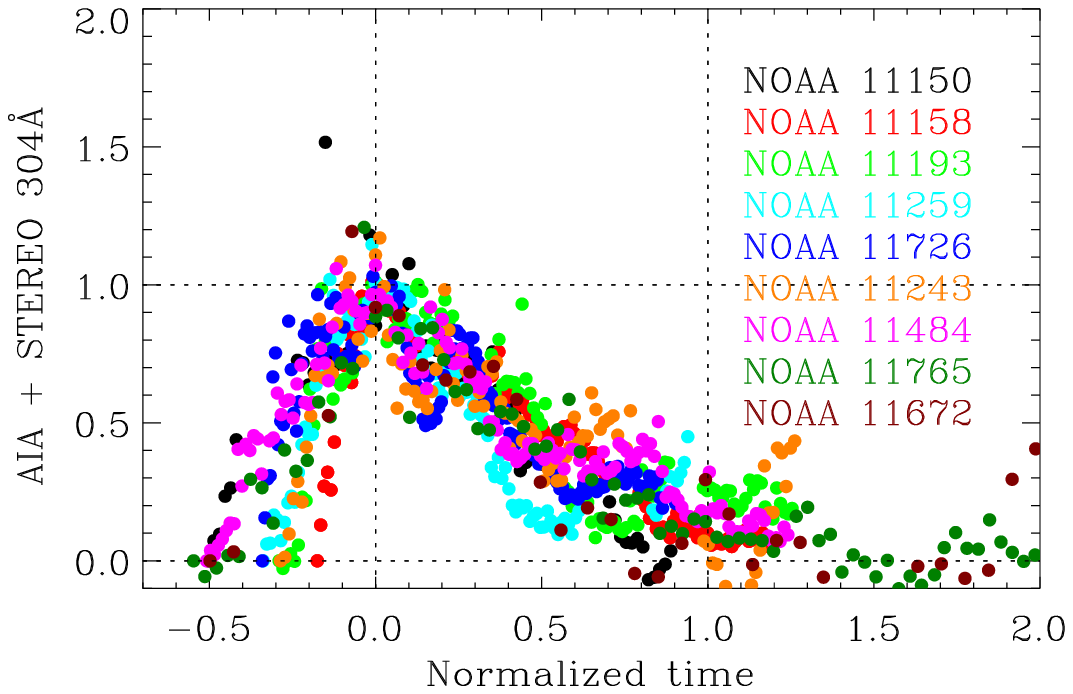}
\caption{304 \AA\ integrated intensity light curves for the full active region data set. Left
  panel: light curves have been shifted in time to align their peak intensities to the origin in
  time. Right panel: light curves have been scaled in intensity to match the peak in NOAA 11158
  and then normalized by that peak intensity. The intensity scaling factor is used to scale the
  times, then normalized by the NOAA 11158 peak-to-disappearance duration.}
\label{fig:ARcurves}
\end{figure*}
\begin{deluxetable}{rrrrr}
\tablecaption{Active region dataset}
\tablewidth{0pt}
\tablehead{
	       & 		& \multicolumn{2}{c}{Observing window} & \\
   Number      & Latitude	& \multicolumn{1}{c}{Start}   & \multicolumn{1}{c}{End} & Cadence   }
\startdata
11150 & -21.5& 2011/01/01	& 2011/02/27	  & 12h \\
11158 & -21.5& 2011/02/01	& 2011/04/18	  & 12h \\
11193 &  17.1& 2011/04/04	& 2011/06/14	  & 12h \\
11243 &  15.8& 2011/06/11	& 2011/07/25	  & 6h  \\
11259 &  26.4& 2011/07/05	& 2011/08/13	  & 6h  \\
11272 & -13.0& 2011/07/28	& 2011/09/16	  & 6h  \\
11484 &  12.0& 2012/05/07	& 2012/07/22	  & 12h \\
11672 & -19.3& 2013/02/05	& 2013/02/29	  & 12h \\
11726 &  15.0& 2013/04/14	& 2013/07/07	  & 12h \\
11765 &   7.9& 2013/05/31	& 2013/07/27	  & 12h
\enddata
\label{tab:dataset}
\end{deluxetable}

\section{Observations}

The STEREO Ahead (A) and Behind (B) spacecraft are, respectively, leading and trailing the Earth
in its orbit around the Sun, providing multiple observational vantage points. We used
304~\AA\ images from EUVI/STEREO \citep{howard2008} and AIA/SDO \citep{lemen2012} to study the
continuous evolution of active regions during their complete lifetimes.  For the data set
presented here, the separation angle between the two spacecraft and SDO (with the Earth
perspective) ranges between $80-140\degr$, providing a complete look at the solar surface visible
from the ecliptic plane.

The data set (Table~\ref{tab:dataset}) consists of 10 active regions manually selected. The
criterion for selection was that the evolutionary path (from emergence to growth and maximum
development, and to complete decay) was minimally impacted by neighboring active regions. While
this is not the most common evolutionary scenario for an active region (because interactions with
other regions are frequent) it allows us to study the elemental processes at play in magnetic flux
evolution. Still, regions such as 11726, one of the longest-lived, do interact with smaller shorter lived
flux concentrations.

To track the history of each individual active region, we made full-Sun Heliographic maps from
near-simultaneous 304 \AA\ EUVI (A and B) and AIA images. We chose solar rotation at the latitude
of the active region, which conveniently maintains the longitudinal position of the region. Images
were processed using standard software provided by the instrument teams.  AIA 304 \AA\ images were
scaled to EUVI intensities with a time dependent factor available at the SolarSoft STEREO beacon
directories. This correction is estimated from the ratio of the median on-disk brightness of EUVI
(average of A and B) and AIA hourly images. It is not meant as an absolute calibration correction,
but it does remove the degradation in the AIA 304 \AA\ response with respect to EUVI. The maps
were made at a spatial resolution of $1\degr$ in longitude and latitude.

Line-of-sight HMI/SDO \citep{Scherrer_etal12, schou2012} magnetograms were used to compute the
total unsigned flux of the active regions as a function of time during their pass along the Earth
side view. Both the pixel areas and the observed flux densities were
corrected for the line-of-sight projection angle \citep[see][]{hagenaar2001}.

\section{Results}
\subsection{Active region evolution curves}

We characterize the evolution of the active regions using 304 \AA\ light curves, which are good
indicators of the general trend of emergence, growth, and decay in an active region
\citep[e.g.][]{ugarte-urra2012}.  The light curves are calculated by integrating the excess counts
from a background image in all pixels within a $21\degr\times21\degr$ field-of-view around
the center of each region in the Heliographic maps. The background image, which approximates the background of the
quiet Sun pre-existent to the active region emergence, is estimated from the median intensity of
the bottom 20\% intensities in time at each pixel, including times before the emergence. Data is discarded if it lies beyond 0.9
R$_{\sun}$ of the center of the EUVI and AIA images to avoid the large interpolation edge effects
near the solar limb. Elsewhere a correction is implemented based on a parameterization of the
dependence of these effects with longitude at the latitude of interest
\citep[see][]{ugarte-urra2012}.

\begin{figure}[htbp!]
\centering
\includegraphics[width=\columnwidth]{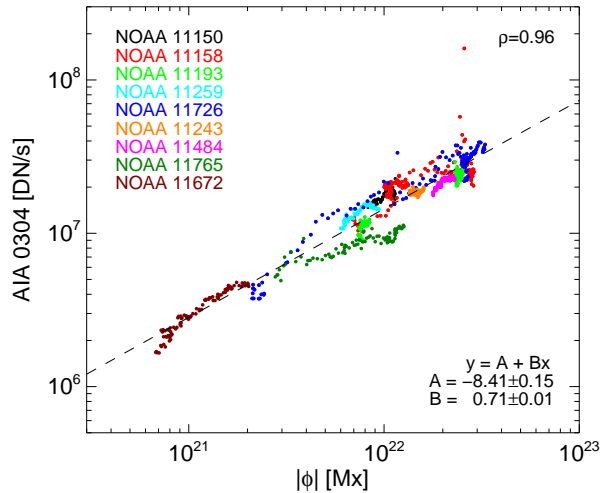}
\caption{Scatter plot of AIA 304\AA\ counts/s vs. total unsigned magnetic flux.}
\label{fig:flux-lum}
\end{figure}

The left panel in Figure~\ref{fig:ARcurves} shows the light curves of the active regions in the
data set with the times of peak in 304 \AA\ intensity co-aligned. The times of peak intensity were
determined by smoothing the light curves with a 24h time window, thus minimizing spikes due to
transient activity like flares. To take into account different quiet Sun background levels, we
have also subtracted the starting intensity from each light curve. The curves reflect the well
known evolution of active region magnetic fields with emergence leading to a rapid rising phase
followed by a longer decay period.  In bipolar active regions, the rise time to maximum
development (when emergence ceases) is typically shorter than five days \citep{harvey1993}, but it
can be longer if multiple bipoles are involved and this is the case for some regions in our data
set. The decay in most regions in our sample is close to monotonic, but there are regions, e.g.,
11726, that exhibit large variations in 304 \AA\ intensity which are due to
interaction with shorter lived nearby flux.

Also evident is that the brighter regions live longer. This is reminiscent of the observed
property that active regions with larger fluxes and areas have longer lifetimes
\citep[e.g.][]{mosher1977}. As one will expect intuitively (and we will show quantitatively later
in the paper) this is not coincidental, but it has its foundation in the relationship between
magnetic flux and 304 \AA\ intensity. An interesting property, that to our knowledge has not been
reported before is that the evolution light curves are scalable. In other words, by knowing the
peak intensity of a region in 304 \AA, one can estimate its lifetime. This is demonstrated on the
right panel of Figure~\ref{fig:ARcurves}. In this diagram we show all the light curves scaled to
the active region 11158 peak intensity. That same scaling factor is used to scale the time,
meaning that an active region twice as bright at peak, will live twice as long. In the figure,
intensities have been also normalized by the AR 11158 peak intensity, and the times by the 11158
lifetime, set as the time when the 304 \AA\ intensities reach the flat quiet Sun levels marginally
above the original quiet Sun. That plateau is discussed by \citet{sainzdalda2008}.

To understand the intensity decay, it is important to understand the relationship between the 304
\AA\ maps and the magnetic field because it is the magnetic field evolution on the surface that
drives the upper atmosphere response.

\subsection{Magnetic flux from 304 \AA\ images}

The relationship between the X-ray and EUV intensities and magnetic quantities, such as the total
unsigned magnetic flux in an active region, has been firmly established
\citep[see][]{gurman1974,schrijver1987,fisher1998,pevtsov2003,vandriel2003,fludra2008}.
Chromospheric EUV lines like \ion{He}{1} 584.34 \AA\ are no exception, as shown by CDS/SoHO
measurements \citep{fludra2002,fludra2008}. We show in Figure~\ref{fig:flux-lum} that this
relationship also holds true for \ion{He}{2} 303.8 \AA, the dominant spectral line in the 304
\AA\ channel \citep{odwyer2010}.

In Figure~\ref{fig:flux-lum} we plot the total number of AIA 304 \AA\ counts per second in an active region as a
function of total unsigned flux as measured from HMI magnetograms. The data points represent
observations tracking each active region from limb to limb with a cadence of 1 hour. AIA 304
\AA\ images were corrected for the time dependent degradation in sensitivity in the filter
response, provided by the instrument team within the standard processing software.
A threshold of 85 DN/s/pixel was used to discriminate between the pixels with active region contribution
from the quiet Sun pixels. The field-of view ranged between $215\arcsec-325\arcsec$ depending on the size
of the region, with the exception of active region 11672 in which we used a $100\arcsec\times110\arcsec$ area.
In the magnetograms, only field strengths above 20 Gauss, in absolute value, were considered.

The power law relationship can be expressed as $log(I_{AIA304}) = A + B~log|\phi|$, where
$I_{AIA304}$ is in DN/s, $B=0.71\pm0.01$ and $A=-8.41\pm0.15$. This link between the 304 \AA\ integrated
intensity and the magnetic flux is important, not only for its implications in terms of how the
EUV emission is generated, as the previous studies have argued, but also because
it allows the 304 \AA\ intensity to be used as a proxy of magnetic flux. This is particularly
interesting now, as we do not currently have magnetic field measurements from outside the Earth's
line-of-sight perspective, but we do have 304 \AA\ measurements thanks to STEREO. The 304 \AA\ emission is better
than other lines at tracking flux evolution. It is formed closer to the solar surface than other
EUV and X-ray lines, which extend over larger volumes in the atmosphere and are more sensitive to
transient energetic events that generate significant density changes with their respective
intensity fluctuations.

This power law can be used to translate the 304 \AA\ light curves in Figure~\ref{fig:ARcurves} into
total unsigned magnetic flux curves for the full lifetime of the active regions. An intermediate
step is necessary to go from the integrated intensities of the STEREO heliographic maps to the AIA
intensities in the powerlaw. That relationship in our analysis is: $log(I_{AIA304}) = C +
D~log(I_{HG304})$, with $C=-0.9\pm0.3$ and $D=1.11\pm0.04$.

To asses the importance of the chosen thresholds in the flux-luminosity relationship we repeated the
analysis for all regions in different scenarios: AIA thresholds of 5, 85 and 190 DN/s and HMI thresholds
of 20, 40 and 80 Gauss. While the slopes of the power law can change significantly in all those
permutations due to the added or reduced 304 \AA\ counts and magnetic fluxes, the correlation coefficients
are larger than 0.90 in all of them. In the case we selected is $\rho=$0.96. The best fits correspond to an AIA
threshold of 85 DN/s with marginal differences in the correlation and fit uncertainties between the
20 (our choice), 40 and 80 Gauss cases. The fit coefficients for the 40 and 80 Gauss cases are respectively
$A=-7.69$, $B=0.68$ and $A=-6.83$, $B=0.64$.

\section{The Advective Flux Transport model}

{\it Surface Flux Transport} (SFT) describes the latter part of the dynamo process
\citep{Babcock61}, in which the flows on the surface of the Sun transport the magnetic flux from
the active region belts to the poles, canceling the previous polar fields during solar maximum and
then creating a new poloidal field with the opposite polarity. The strength of this new poloidal
field (at solar minimum) is the seed to the next solar cycle \citep{Babcock61,munozjaramillo2013}.

SFT begins with the emergence of bipolar magnetic active regions with a characteristic polarity
and tilt, i.e., with Hale's polarity and Joy's Law tilt \citep{Hale_eta19, Howard1991}. The active
regions emerge at progressively lower latitudes as the cycle progresses, starting  at about $30^{\circ}$
and eventually stopping near the equator \citep{hathaway2011}. The magnetic flux in the active
regions is shredded off by turbulent convective motions (over a period of a few days or weeks) and
is dispersed into the surrounding plasma.

The flux is then transported by the surface flows: differential rotation, meridional circulation,
and the cellular and turbulent motions of convection. The weak magnetic elements are carried to
the boundaries of the convective structures (granules and supergranules) by flows within those
convective structures, forming a magnetic network. Once concentrated in the magnetic network, the
flux is then carried, along with the convective cells, by the axisymmetric differential rotation
and meridional circulation.

While the majority of the active region flux will cancel with the opposite polarity from the
active region itself or future active regions, some residual flux remnants will remain. The lower
latitude leading polarity flux remnants will eventually cancel across the equator, while the
higher latitude following polarity flux remnants migrate to the poles. The following polarity flux
merges with the original global poloidal field and creates a new poloidal field with opposite
polarity, from which the new solar cycle is born.

Most previous SFT models have been highly parametrized, in particular with respect to active
region emergence, the meridional flow, and the convective motions. Previous models have been
restricted to simulating active region emergence by inserting artificial bipolar active region
sources (though some have been based on observed active regions). The adopted meridional flow
profiles (sharply peaked at low latitudes, stopping short of the poles, exaggerated variations
around active regions) deviate substantially from the observed profiles. Additionally, these
models have typically neglected the variability in the meridional flow altogether (the
meridional flow is faster at solar cycle minimum and slower at maximum). Furthermore, virtually
all previous models have parametrized the turbulent convection by a diffusivity with widely
varying values from model to model.

\cite{UptonHathaway14A, UptonHathaway14B} have developed a new surface flux transport model, the
{\it Advective Flux Transport} (AFT) model. This model is built on the SFT foundation created by \cite{DeVore_etal84},
\citet{Sheeley_etal85}, and \citet{Wang_etal89}. However, while early SFT models were used to probe and help
constrain the flows on the Sun, the AFT model was designed with the intent of creating the most
realistic SFT model possible by incorporating the observed active regions and surface flows
directly, with minimal parametrization.

Upton and Hathaway measured the axisymmetric flows by using feature tracking \citep{HathawayRightmire10,
  HathawayRightmire11, RightmireUpton_etal12} on full disk images of the Sun's magnetic field
obtained from space by the MDI/SoHO \citep{Scherrer_etal95} and by the HMI/SDO. The axisymmetric
flows were averaged over each 27-day rotation of the Sun and fit with Legendre polynomials. The
polynomial coefficients were then smoothed in time and integrated into the model.

The AFT model uses a convective simulation to explicitly model the surface flows produced by
the convective flows. The convective simulation, described by \cite{Hathaway_etal10}, uses vector
spherical harmonics to create a convective velocity field that reproduces the spectral
characteristics of the convective flows observed on the Sun. The spectral coefficients evolve,
giving the simulated convective cells finite lifetimes and moving them with the observed differential
rotation and meridional flow. Strong magnetic fields on the Sun inhibit convection; therefore when the
flow velocities are employed, they are dampened where the magnetic field is strong. This magnetic field
strength dependent effect is difficult to reproduce with the diffusivity used in other
models. Advecting the flows with the simulated convection allows the model to surpass the realism
that can be obtained by using a diffusivity coefficient.

Magnetic sources can be incorporated in two different ways: either by manually inserting bipolar
active regions (e.g., using active region databases like NOAA to insert flux daily as the active
region grows) or by assimilating magnetic data directly from magnetograms. This gives the AFT
model additional flexibility. While manual insertion allows the AFT model to be used to
investigate flows and to make predictions, the assimilation process provides the closest contact
to the observations, producing the most accurate syncronic maps of the entire Sun. These maps,
referred to as the {\it Baseline} data set, can be used as a metric for evaluating SFT or as
source data for models that extend into the solar atmosphere and the heliosphere.

\cite{UptonHathaway14A} demonstrated the power of their models by simulating Cycle 23 during the
polar field reversal. Starting with data 3 years ahead of the reversal, they were able to predict
the timing of the field reversal to within four months, with four out of five of their predictions
accurate to within a month. Furthermore, they found that their predictions for the polar field
evolution stayed in remarkably good agreement with the baseline through to the end of the
prediction, an additional 3 years after the reversal, demonstrating the ability to accurately
predict the evolution of the Sun's dipolar magnetic field several years in advance.

\cite{UptonHathaway14B} later went on to use their model to simulate most of Cycle 23 and the
start of cycle 24 (1997-2013) using RGO and NOAA databases to prescribe active region
emergence. They compared the simulated magnetic butterfly diagrams to the baseline and found that
despite limited detail in the active region sources (i.e., total flux, tilt, and longitudinal
separation), the butterfly diagrams were very similar with distinct features often visible in
both. Streams of flux migrating to the poles were also observed, often with a one-to-one
correspondence.

The AFT model is unique in several aspects that make it one of the most sophisticated SFT models to
date. This model produces the most realistic synchronic full-Sun magnetic maps of the Sun
currently available. While the ability of the AFT model to reproduce the evolution of the
global magnetic field has been validated on long time scales, its ability to reproduce fine
details of active region evolution (e.g., active region decay rate) has yet to be investigated. In
fact, \cite{UptonHathaway14B} noted a discrepancy in the amount of flux being injected into the
simulated active region model versus the baseline model. This issue is addressed and resolved in
Section~\ref{sect:magflux_aft}.

\section{Modeling active region flux evolution}
\subsection{Magnetic flux from the AFT model}
\label{sect:magflux_aft}
\begin{figure*}[htbp!]
\centering
\includegraphics[width=\columnwidth]{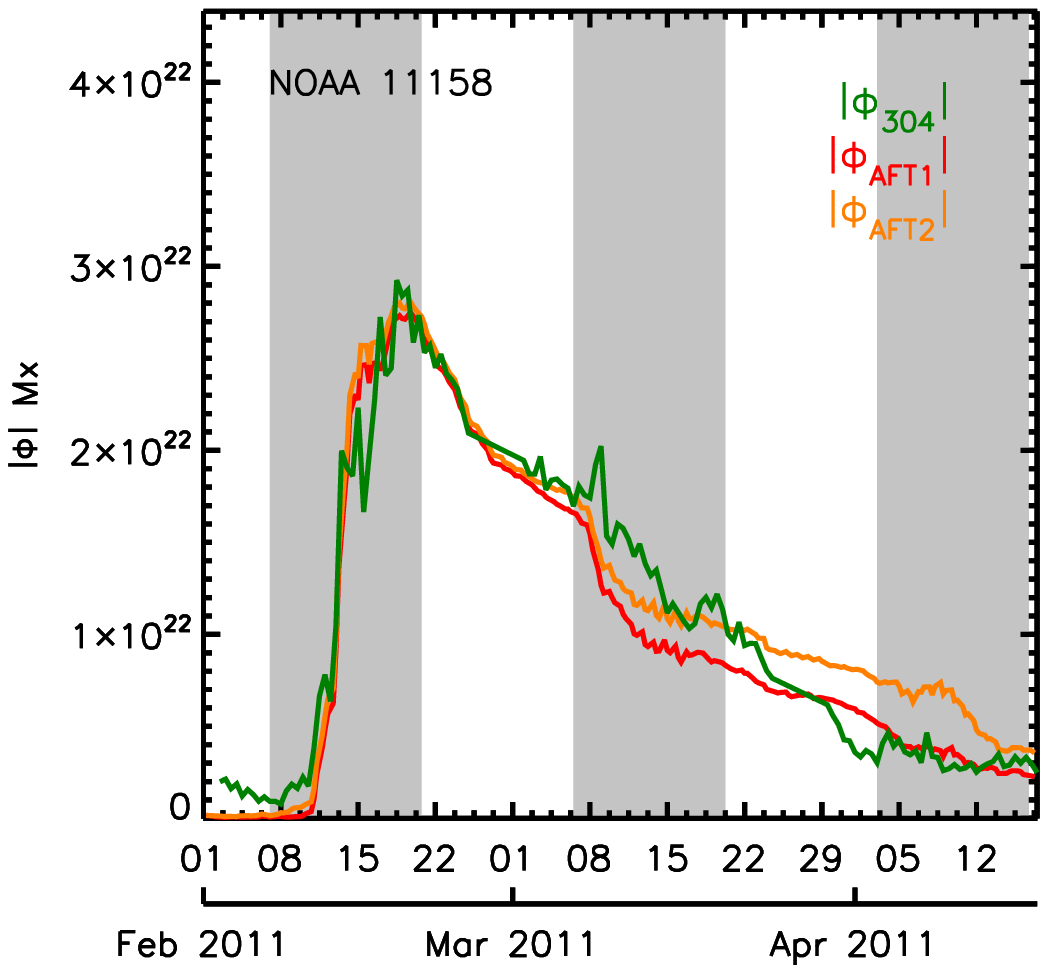}
\includegraphics[width=\columnwidth]{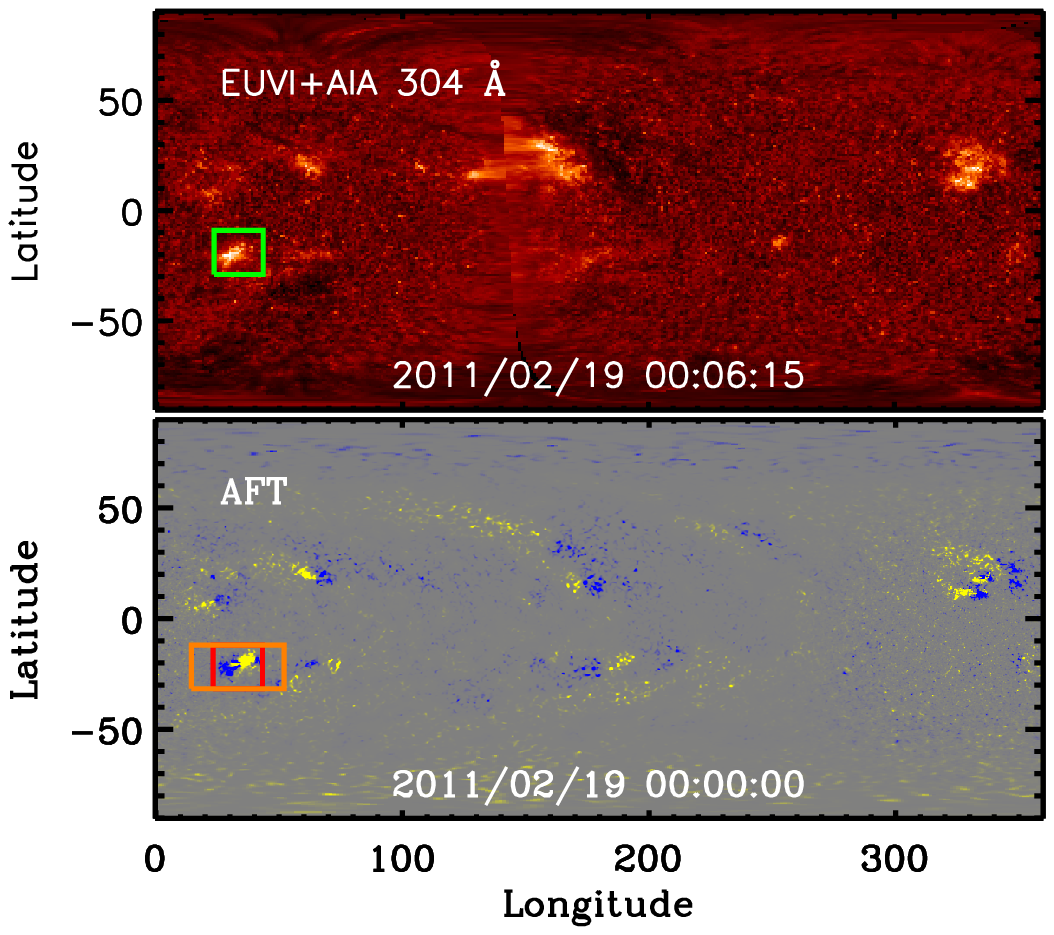}
\includegraphics[width=0.66\columnwidth]{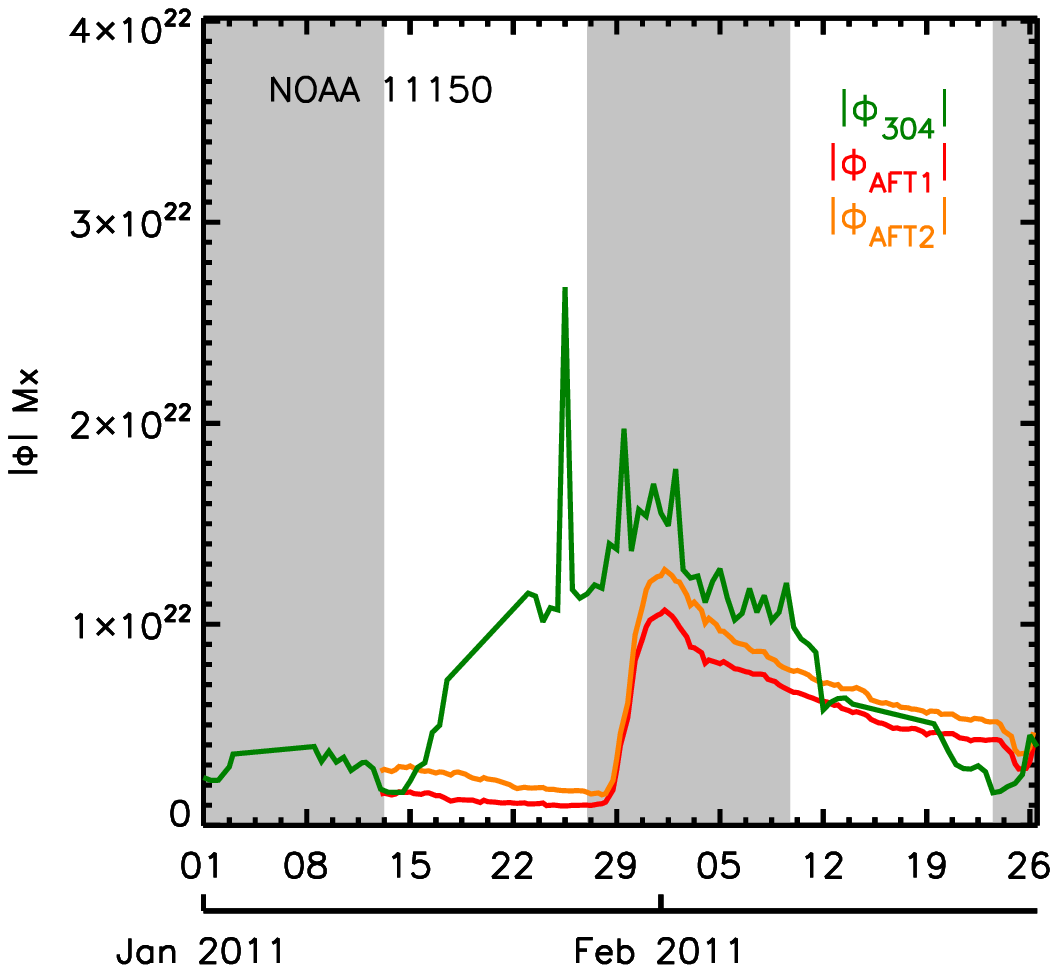}
\includegraphics[width=0.66\columnwidth]{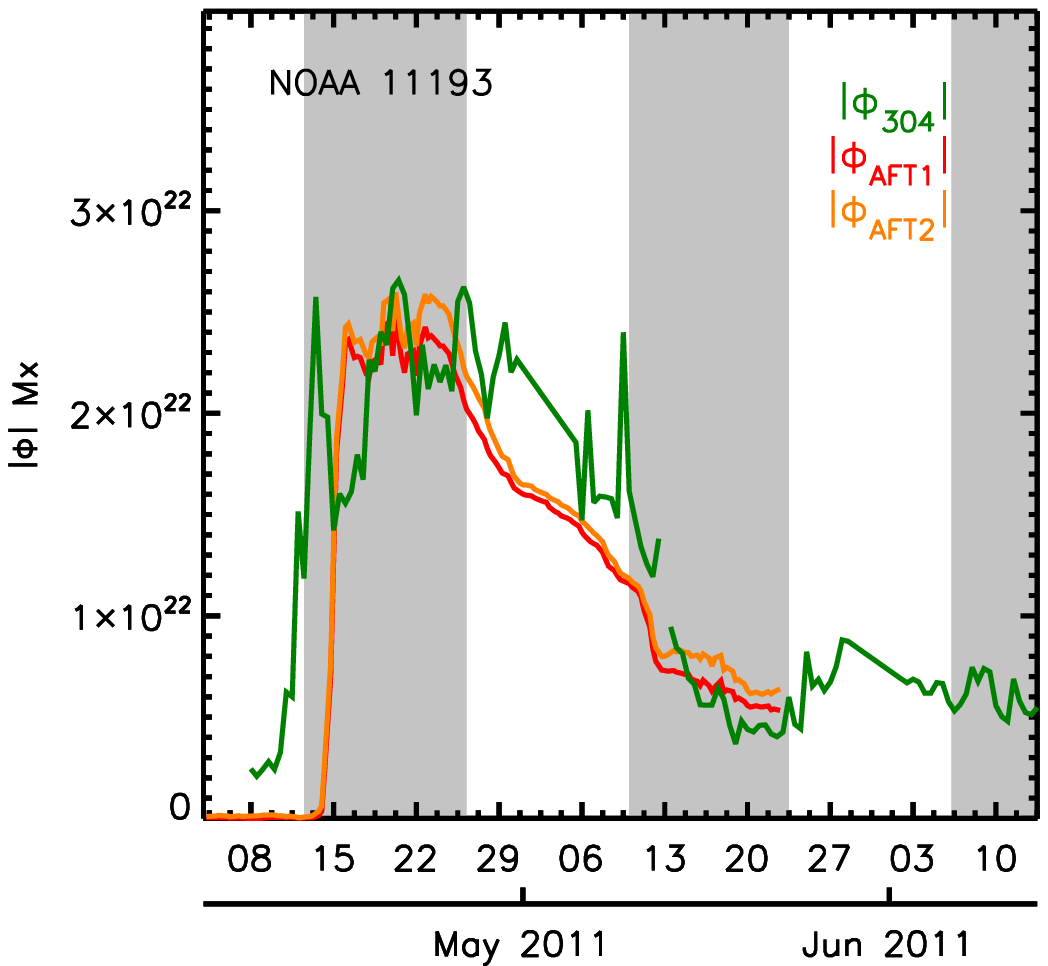}
\includegraphics[width=0.66\columnwidth]{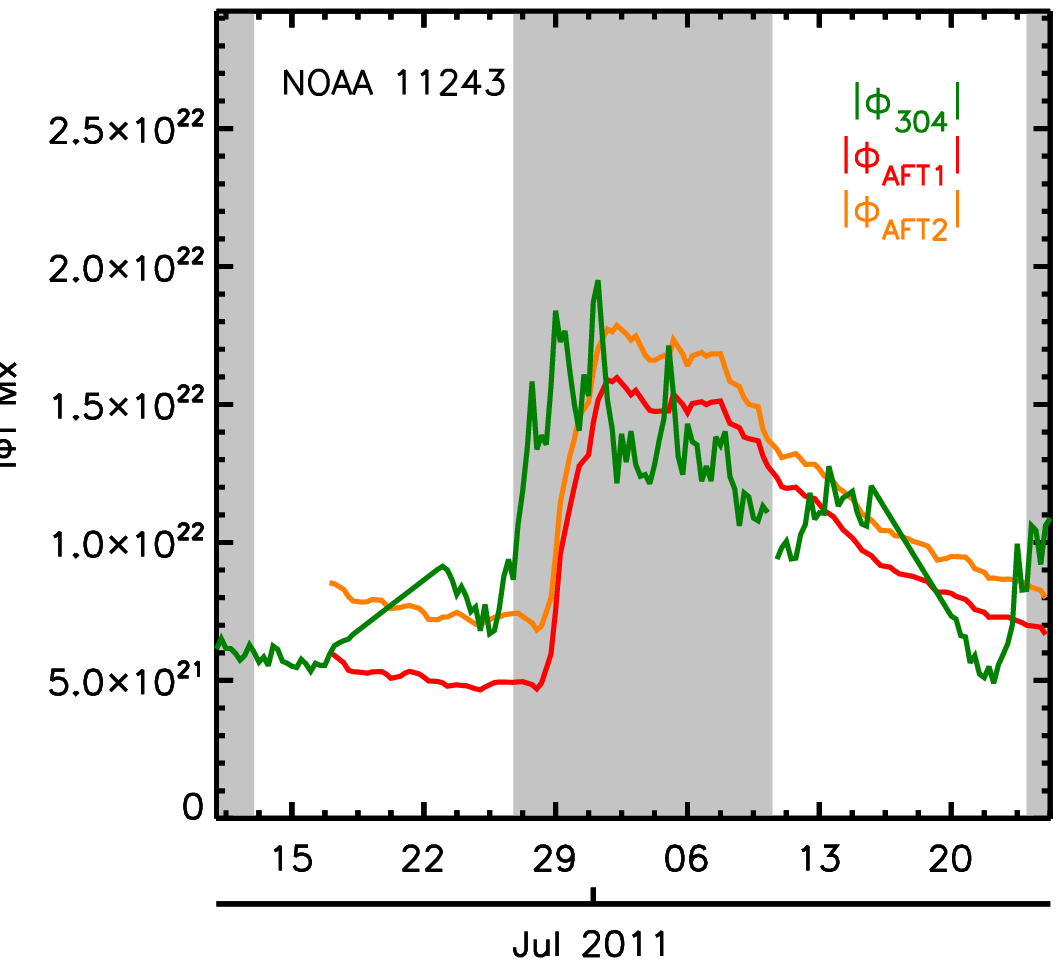}
\includegraphics[width=0.66\columnwidth]{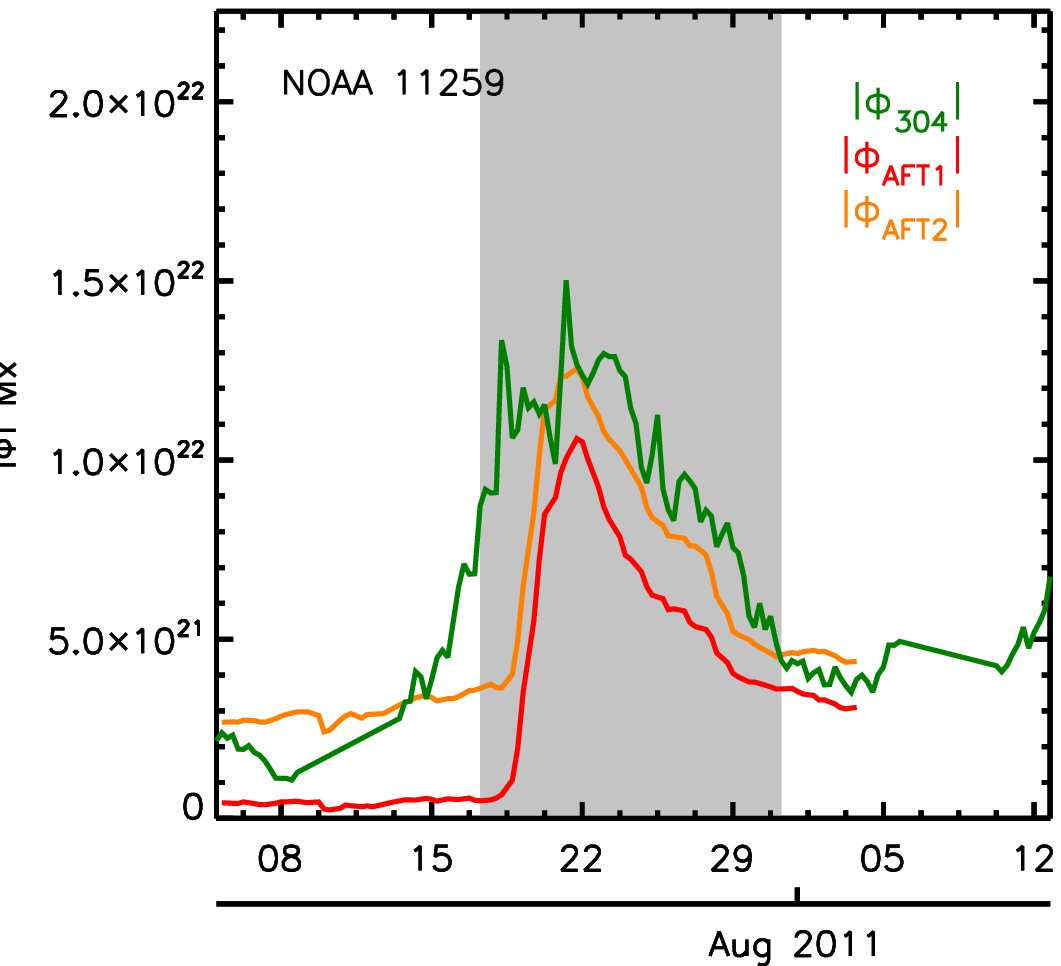}
\includegraphics[width=0.66\columnwidth]{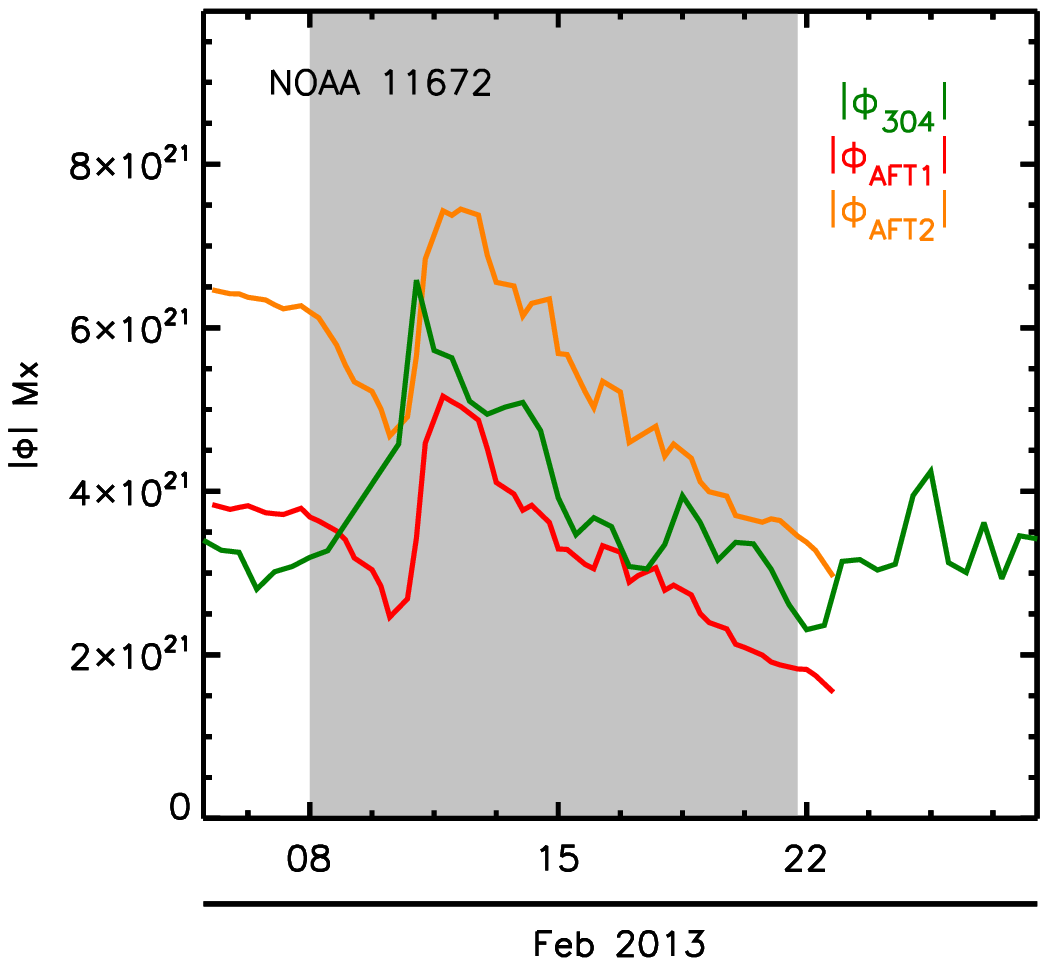}
\includegraphics[width=0.66\columnwidth]{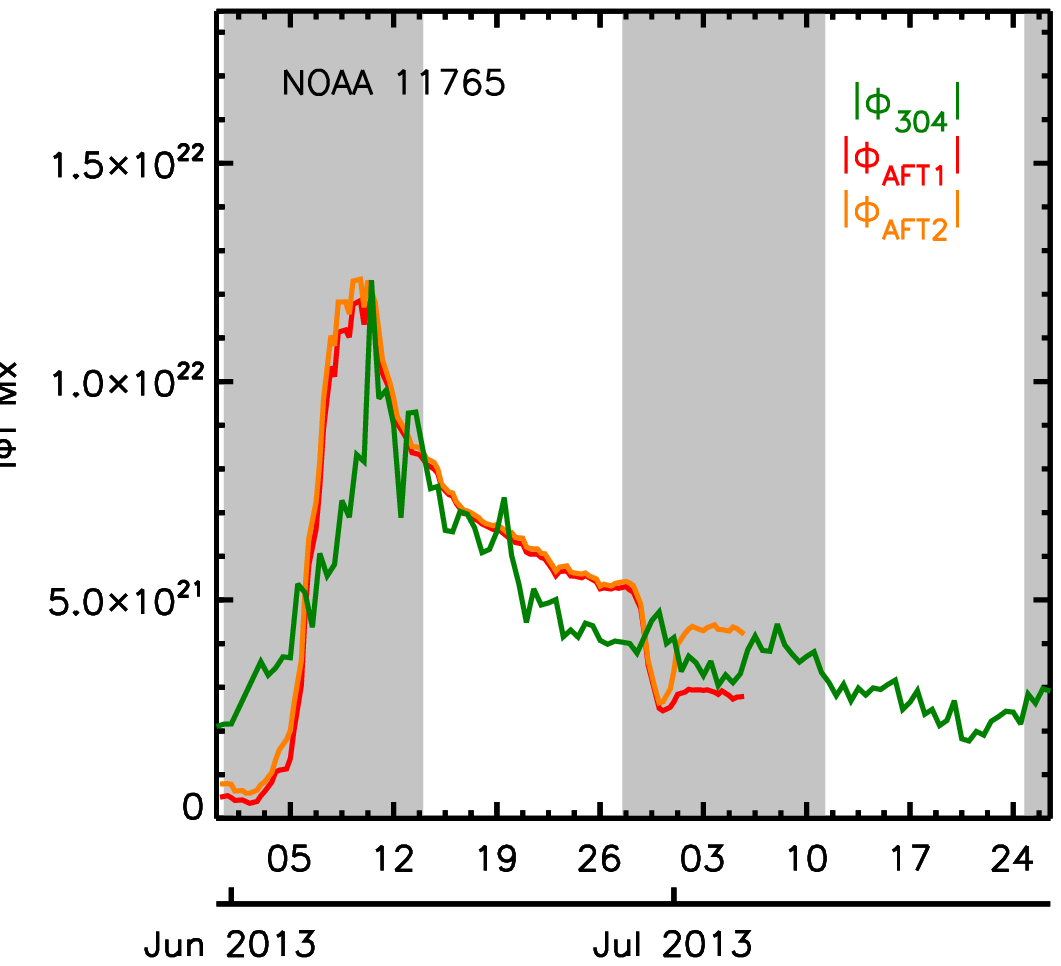}
\includegraphics[width=0.66\columnwidth]{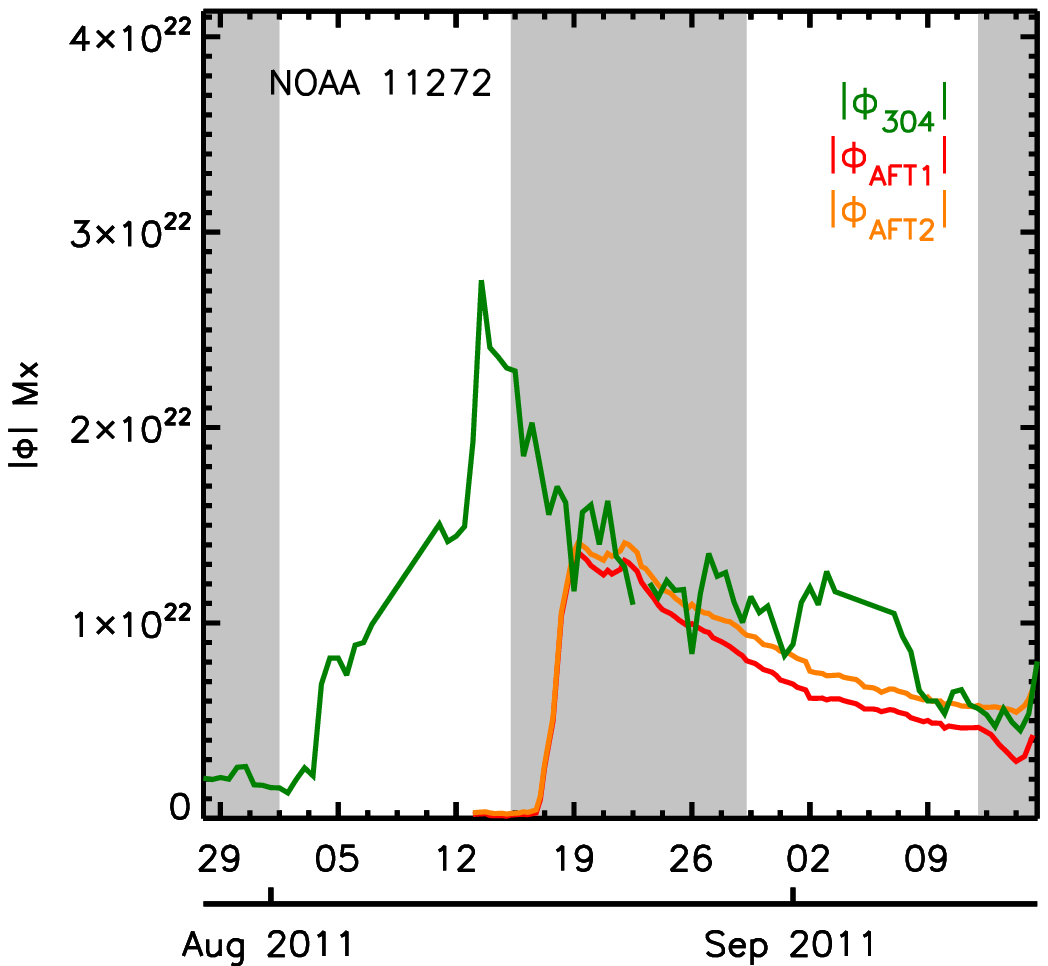}
\includegraphics[width=0.66\columnwidth]{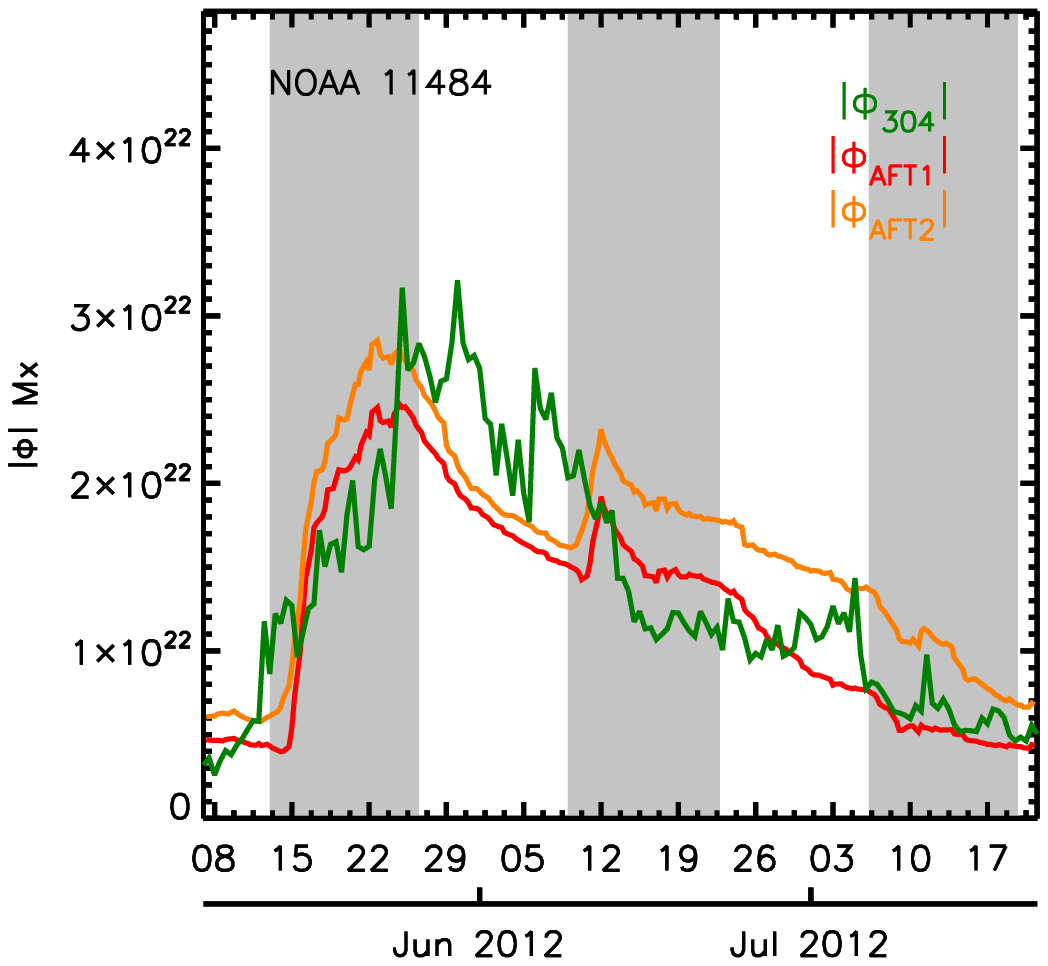}
\includegraphics[width=0.66\columnwidth]{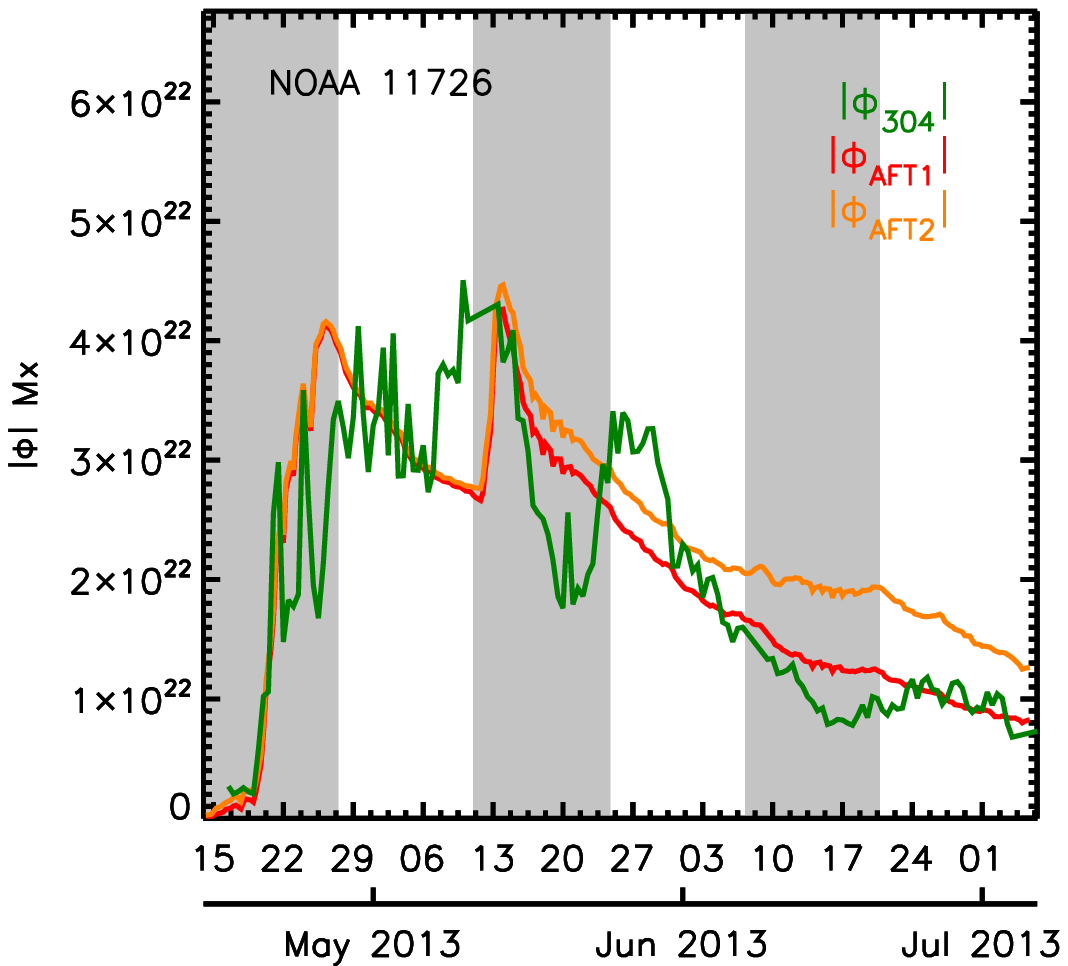}
\caption{Total unsigned magnetic flux of the dataset.  Shown in different colors are the flux
  obtained from the 304 \AA\ proxy (green) and the flux for two area integrations of the AFT
  model (red and orange) in its standard use, that is including assimilation of HMI magnetograms
  (i.e., the baseline). Grey areas mark the times when the active region is on the Earth side,
  when data from HMI magnetograms is being assimilated. 11158 is highlighted on top with the
  corresponding heliographic and AFT maps near peak 304 \AA\ intensity.}
\label{fig:SFTassim}
\end{figure*}

\begin{figure*}[htbp!]
\centering
\includegraphics[width=\columnwidth]{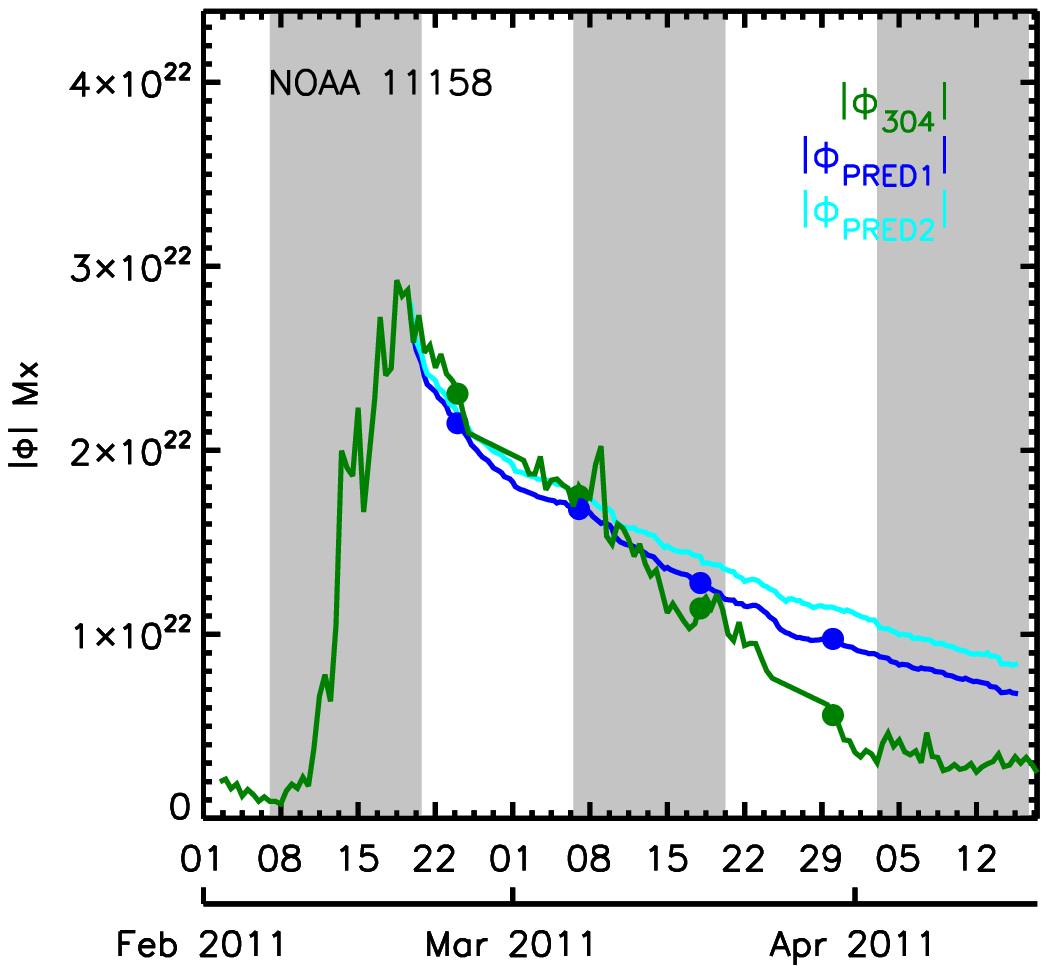}
\includegraphics[width=\columnwidth]{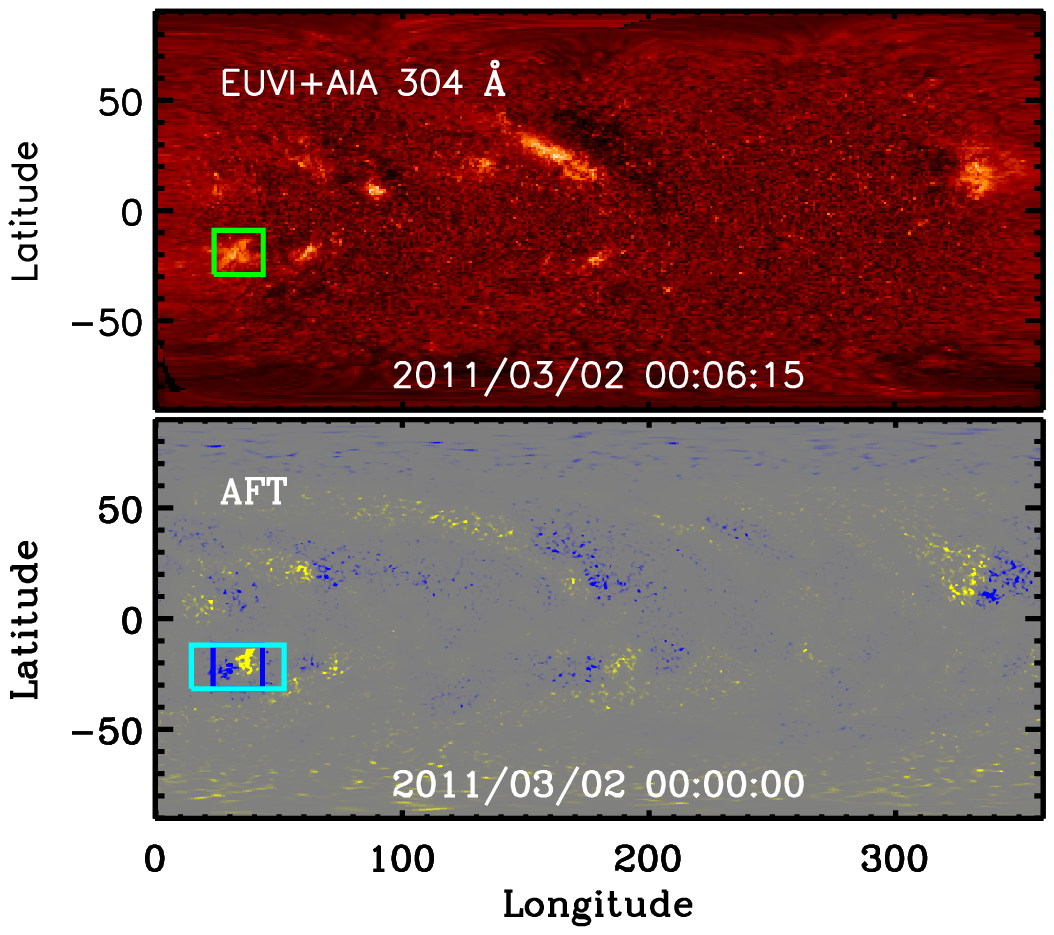}
\includegraphics[width=0.66\columnwidth]{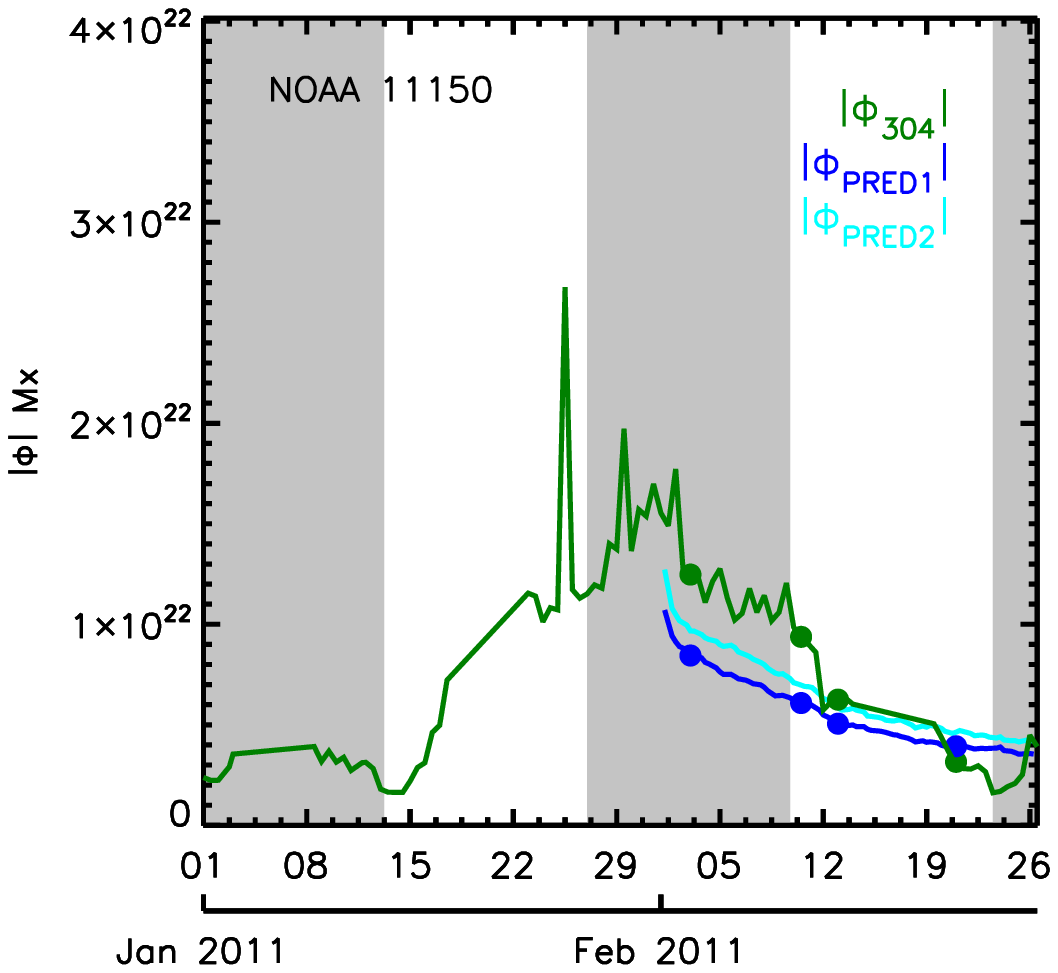}
\includegraphics[width=0.66\columnwidth]{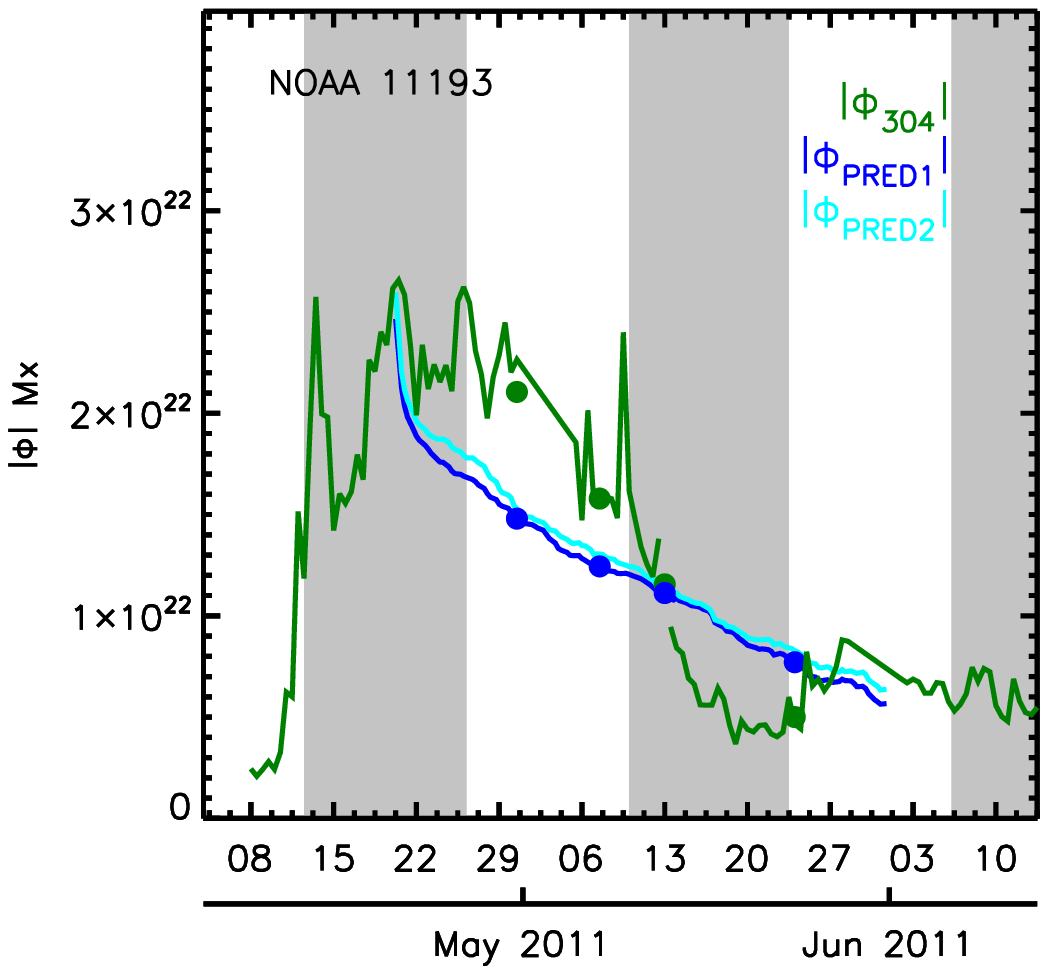}
\includegraphics[width=0.66\columnwidth]{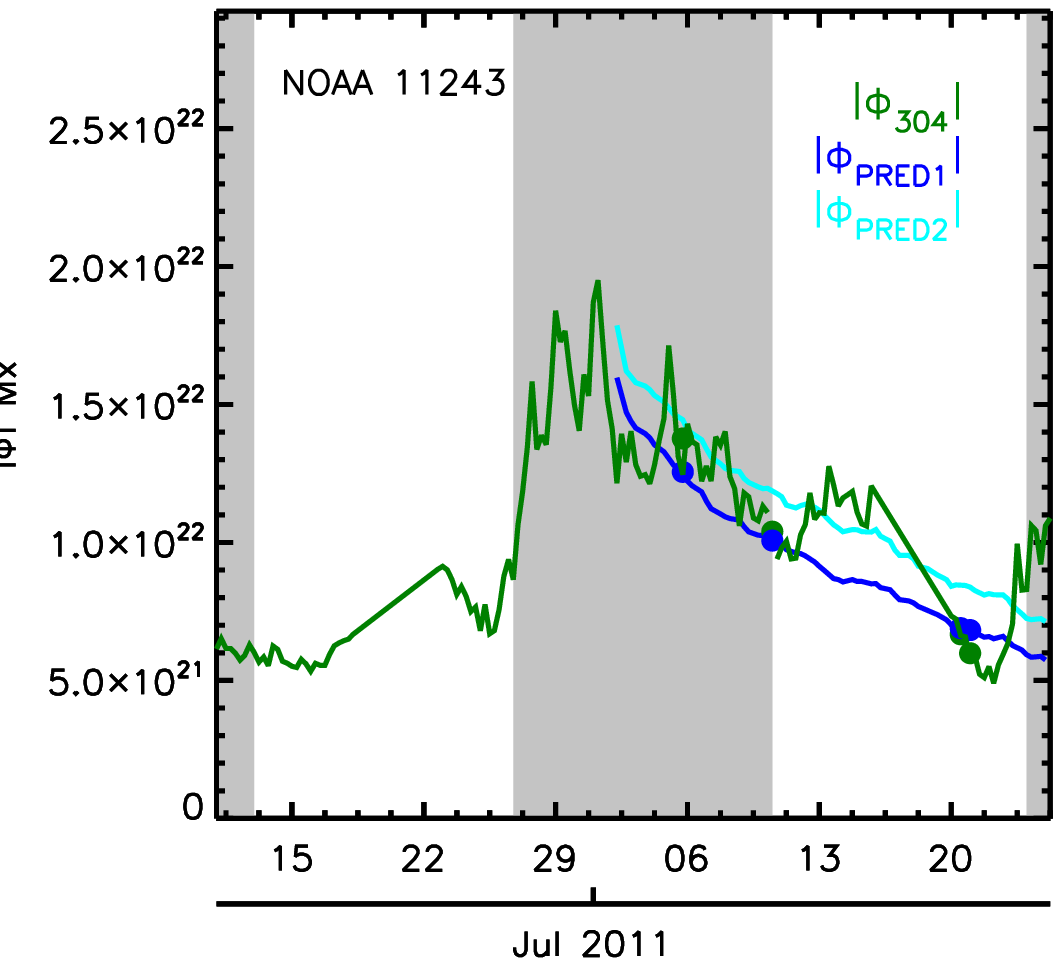}
\includegraphics[width=0.66\columnwidth]{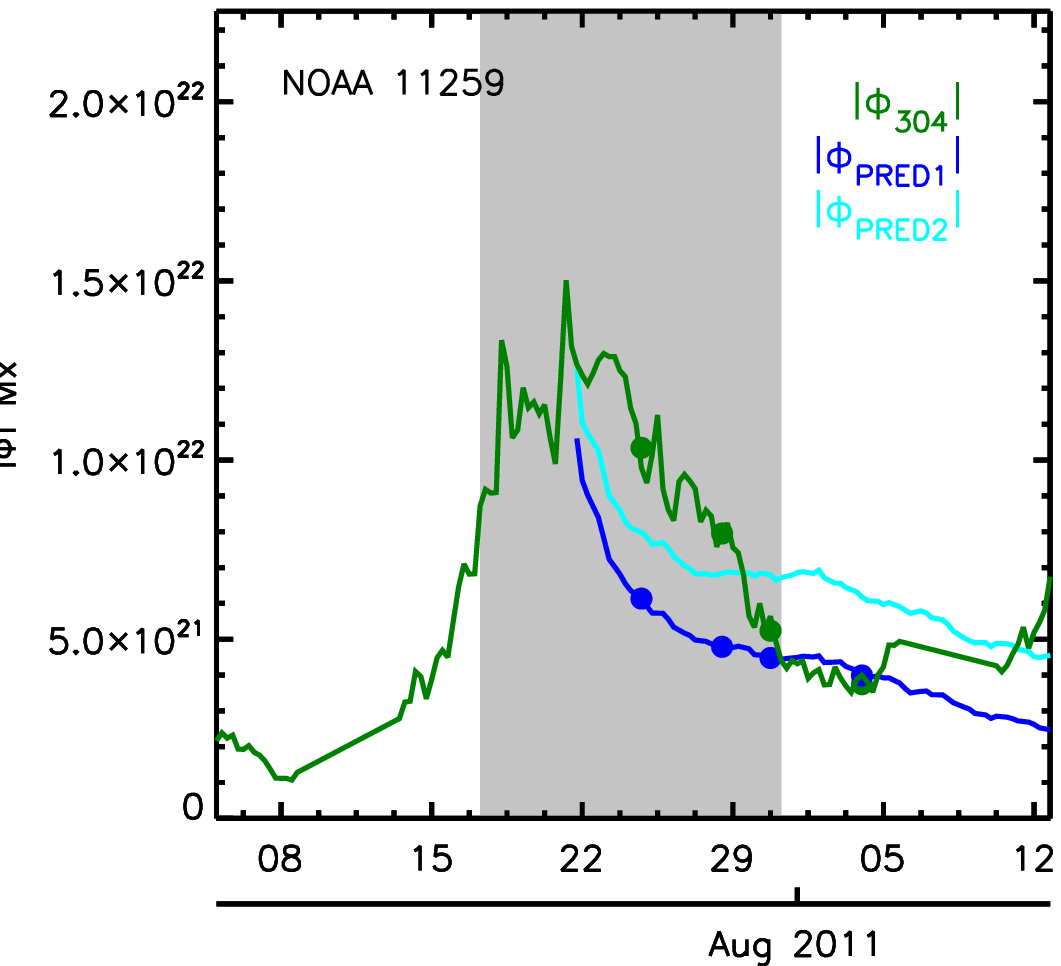}
\includegraphics[width=0.66\columnwidth]{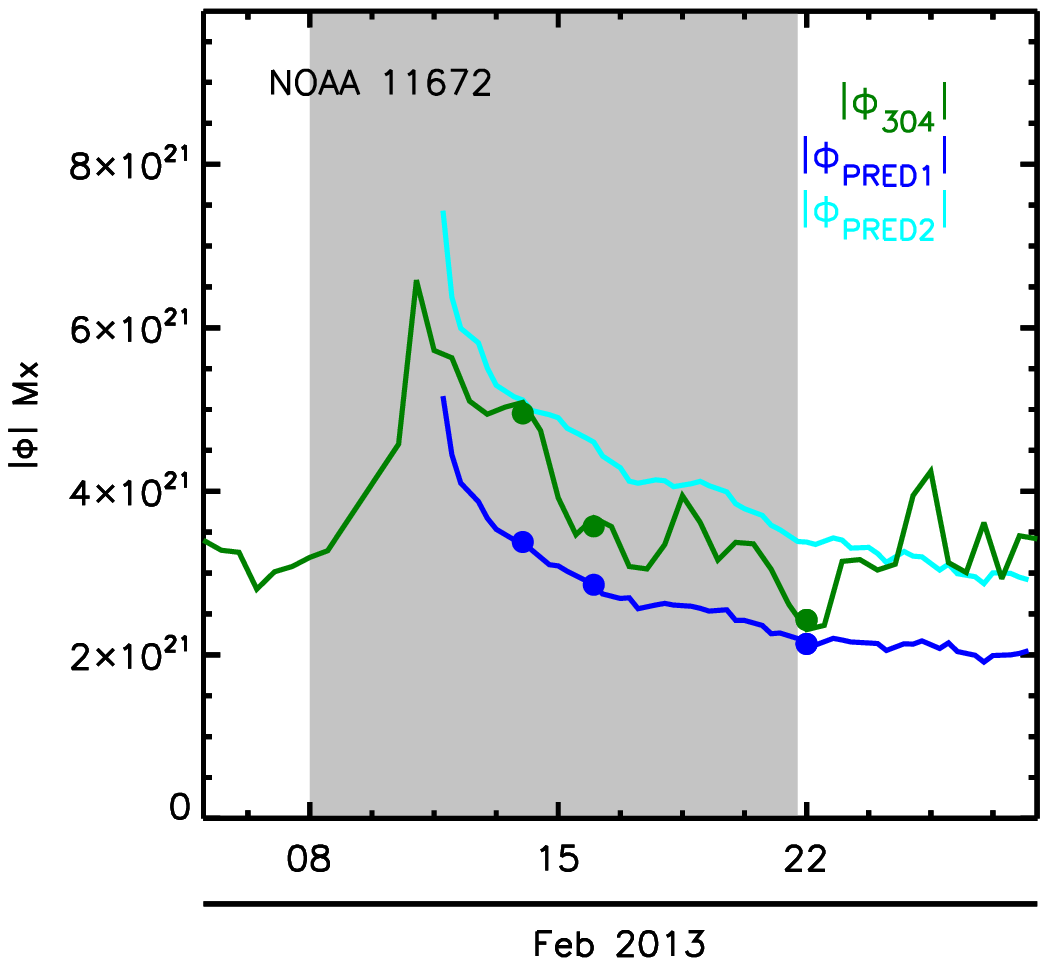}
\includegraphics[width=0.66\columnwidth]{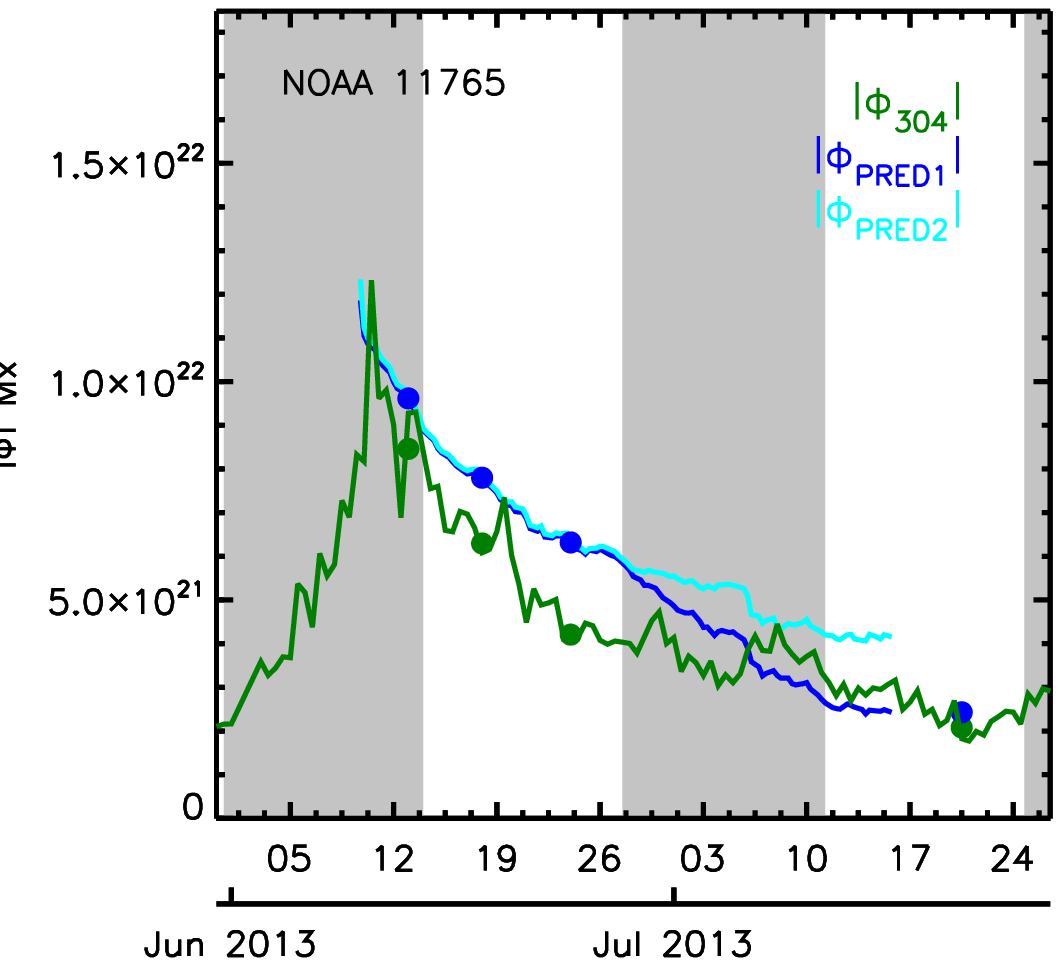}
\includegraphics[width=0.66\columnwidth]{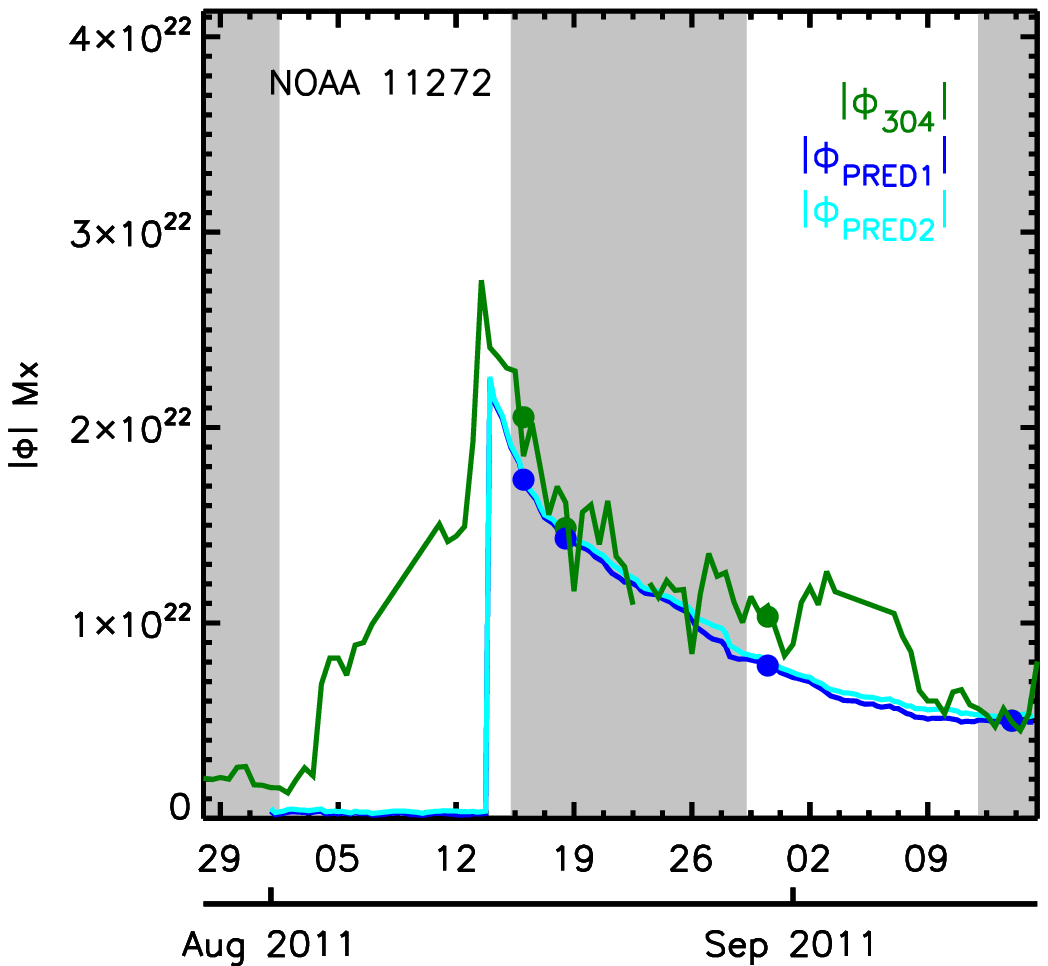}
\includegraphics[width=0.66\columnwidth]{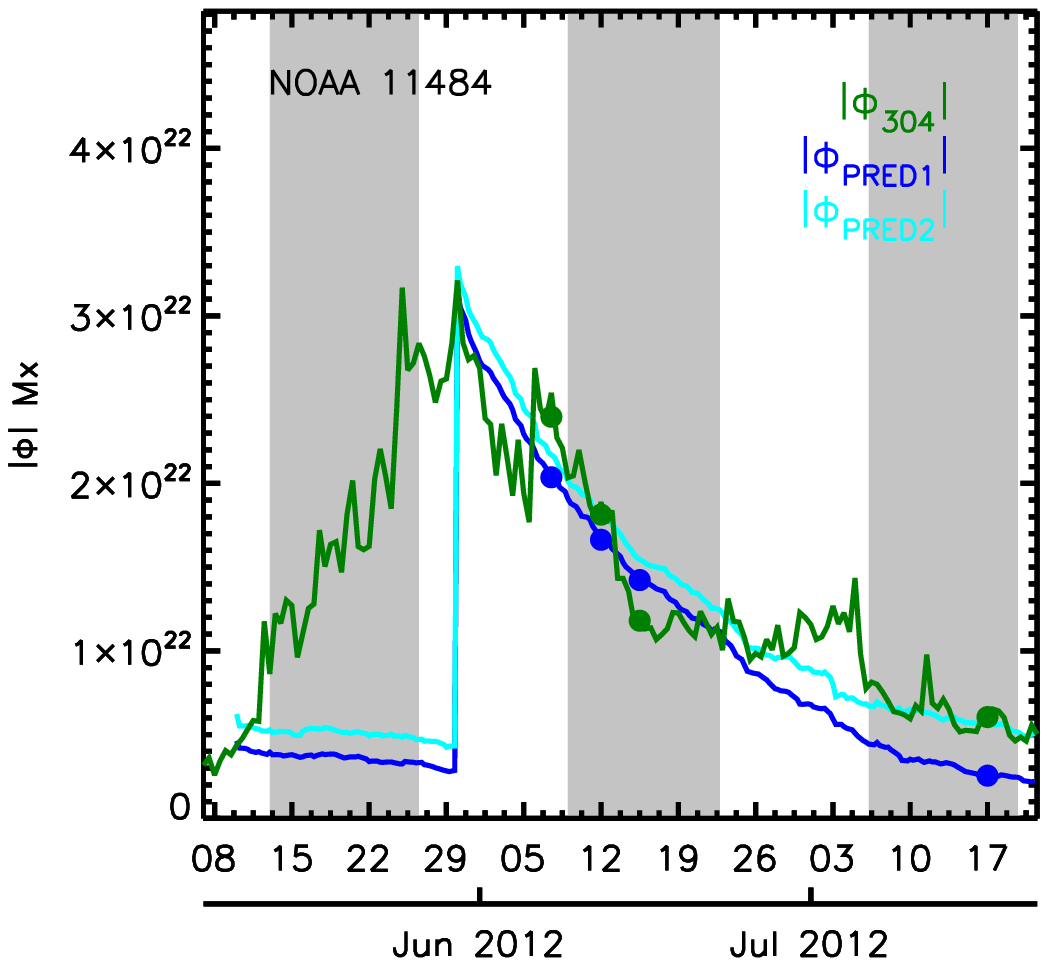}
\includegraphics[width=0.66\columnwidth]{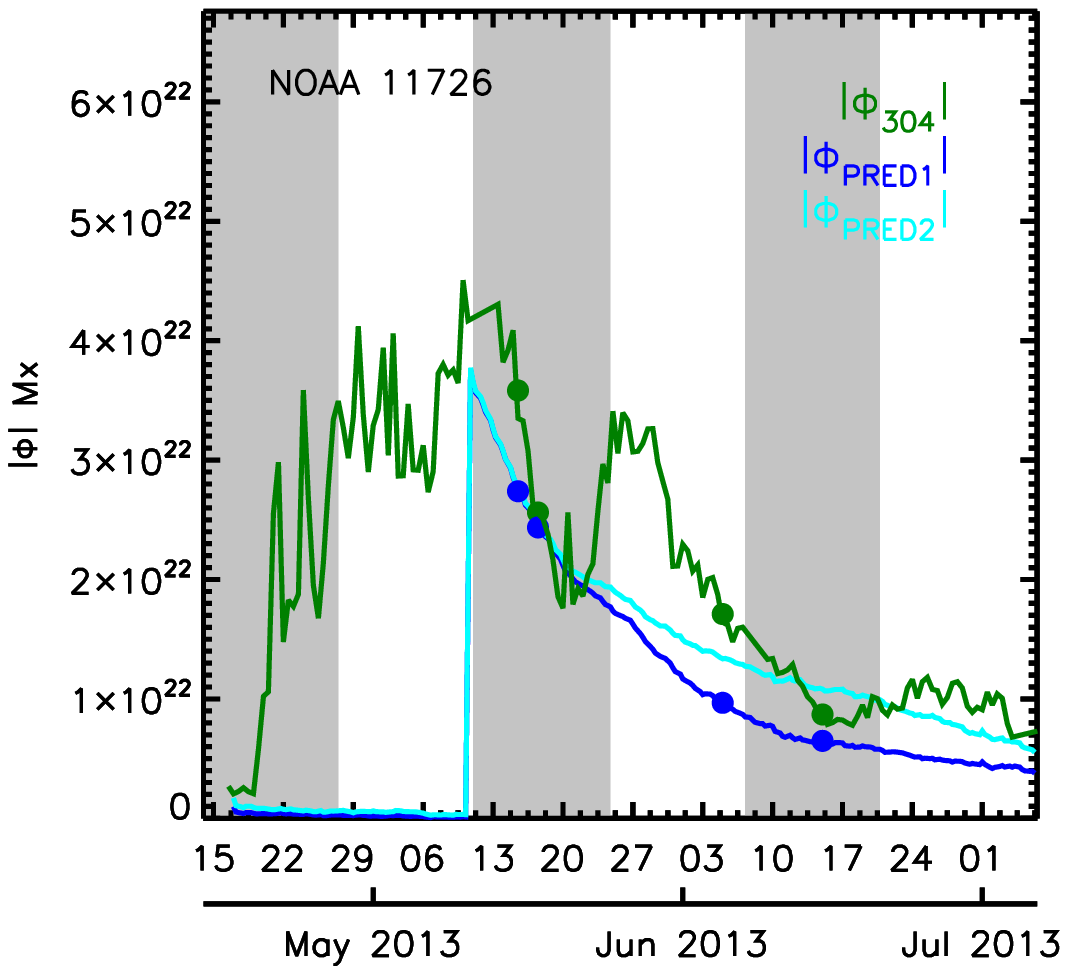}
\caption{Total unsigned magnetic flux as a function of time for the active regions in the dataset.
  Shown in different colors are the flux obtained from the 304 \AA\ proxy (green) and the flux
  decay predictions from the AFT model (blue and cyan) in which assimilation has been stopped
  at the times of peak 304 \AA\ intensity. Grey areas mark the times when the active region is on
  the Earth side. 11158 is highlighted on top with the heliographic and AFT maps well into the
  decay phase. Circles correspond to times when $|\phi_{304}|$ is a fraction (0.8, 0.6, 0.4 and 0.2)
  of its peak, a reference to Figure~\ref{fig:metrics}.}
\label{fig:SFTforecast}
\end{figure*}

While we cannot observe the evolution of the magnetic field when active regions are outside the
field-of-view of available magnetometers such as MDI/SoHO or HMI/SDO (both observing from the
Earth's perspective), there has been significant progress in recent years in simulating how
magnetic field distributions evolve in time. The AFT model is able to evolve observed flux
concentrations by simply applying a velocity field that realistically simulates the flows on the
photosphere, making it possible to compare observed and simulated magnetic
quantities for the complete history of an active region.

\begin{figure*}[htbp!]
\centering
\includegraphics[bb= 0 100 566 300,clip=true,height=4.4cm]{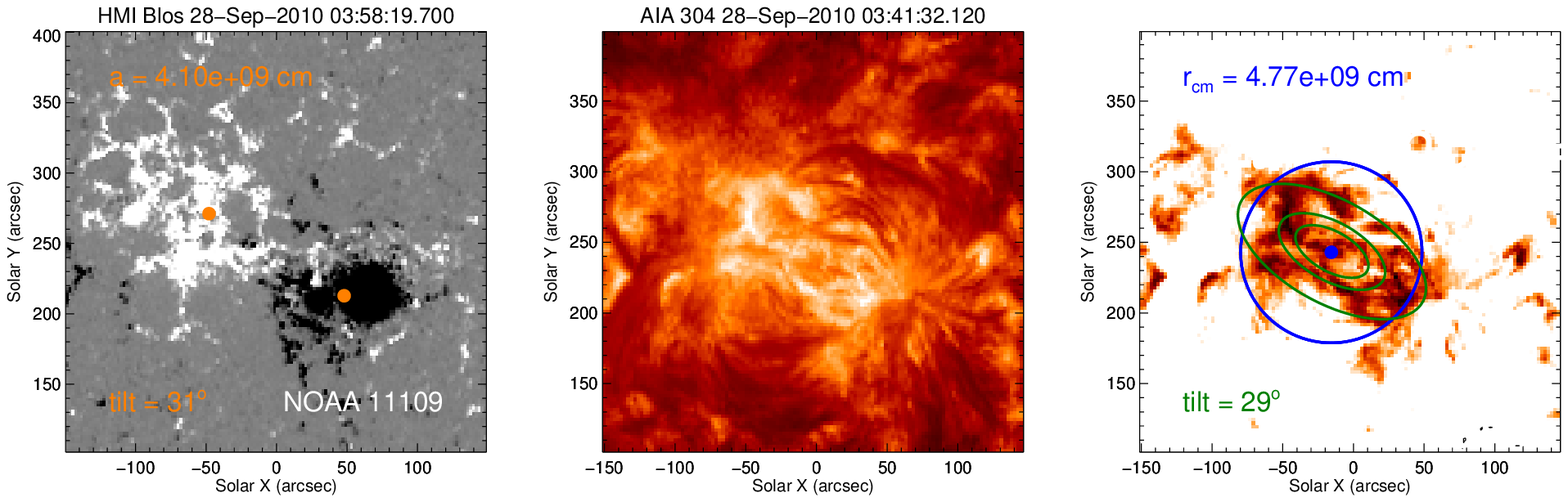}
\includegraphics[height=4.2cm]{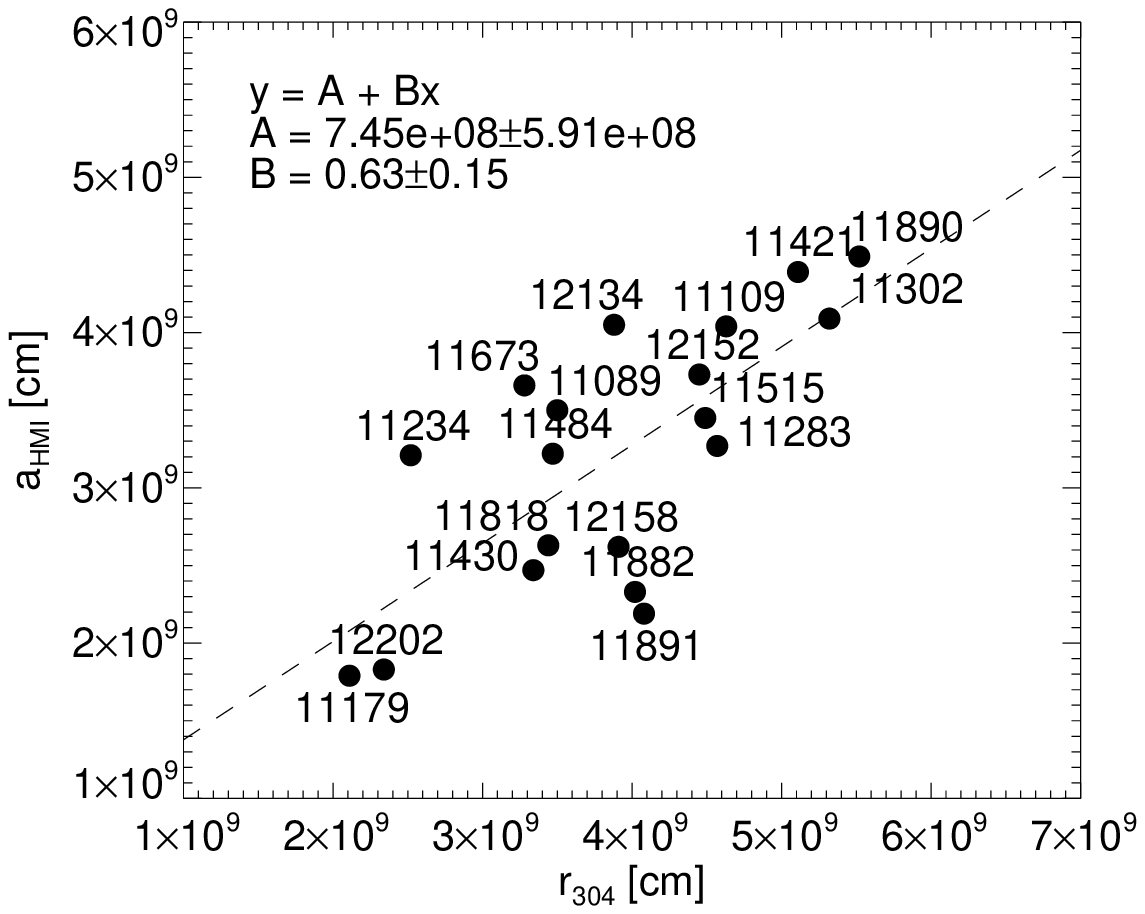}
\caption{The three leftmost panels show a sample case demonstrating the use of 304 \AA\ images to derive properties of the magnetic field distribution.
Left panel: HMI magnetogram with polarity centroids, separation ($2a$) and tilt angle. Middle panel: AIA 304 \AA\ image.
Right panel: AIA 304 \AA\ image with a threshold of 170 DN/s. The blue circle represents the distance computed from the weighted
standard deviation from the 304 \AA\ intensity centroid. The green elipses correspond to a 2D gaussian fit, which in this example
is used to estimate the tilt.
The rightmost graph shows this relationship for 19 active regions near central meridian.}
\label{fig:separation}
\end{figure*}

The AFT baseline model assimilates magnetic data from HMI magnetograms while the region is on
the Earth side. Elsewhere it relies on the prescribed flows to allow the flux concentrations to
evolve in time: moving, merging and canceling with other flux concentrations. We investigated the
details of the active region evolution in the AFT model by comparing the total unsigned
magnetic flux from HMI directly to the total unsigned magnetic flux from the AFT baseline model. It
was initially found that the AFT model underestimated the total unsigned magnetic flux, in
particular for active regions newly emerged on the western hemisphere. This may explain the
flux discrepancy previously noted by \cite{UptonHathaway14B}. These comparisons allowed us to make
improvements to the original AFT baseline by optimizing the weighting used in the assimilation
process such that the baseline total unsigned magnetic flux best matched the observed data, while
continuing to mitigate the limb effects seen in the HMI data.

Figure~\ref{fig:SFTassim} shows comparisons of the total unsigned magnetic flux $|\phi_{304}|$
inferred from the 304 \AA\ light curves, as described earlier, to the one predicted by the AFT
model $|\phi_{SFT}|$ after integrating the flux density in the SFT maps for absolute values larger
than 20 Gauss. The assimilated data is shown by the gray shaded areas in the Figure, while areas
with the white background are being evolved solely by the AFT model. The close match of the
curves during times of assimilation serves as an assurance that our 304 \AA\ proxy approach works
because the AFT model is nearly identical to the real HMI data during this time.

We looked at SFT light curves for two different integration areas, shown with different colors in
Figure~\ref{fig:SFTassim}, one matching the field-of-view of the integration area in 304
\AA\ maps, the second one extending that area 25$\degr$ to the East and West in order to assess
the influence of differential rotation on the comparisons. While making the heliographic maps we
tuned the rotation rate to the latitude of the region; the maps produced by the SFT model are made
at the Carrington rotation rate, thus active regions can rotate out of a fixed longitude. The
comparisons reveal that rotation of flux out of the box is not a major source of
error. For NOAA 11672, the weakest region, enlarging the area increases the flux because the
additional background quiet Sun flux has more weight than in stronger regions.

As illustrated by the 11158 curve, the AFT model can accurately reproduce the active region
evolution when the active region is not being observed by HMI and the model is operating without
assimilation. Additional flux may emerge, however, while the active region is on the far-side
(e.g. 11150, 11272 and 11726). The AFT assimilation process corrects for this new emergence
when this flux is observed by HMI and rotates over the East limb. In that respect, the 304
\AA\ proxy is providing information not available to the model.

\subsection{Forecasting magnetic flux evolution}

In the previous sections we established that the peak integrated intensity in 304 \AA\ lightcurves
can accurately determine the duration of the decay phase in an active region. We also showed that
304 \AA\ can provide information about magnetic properties of an active region when magnetic field
measurements are not available. Finally and complementary to the others, we have seen that the
AFT model has the ability to make these properties evolve realistically in time. The next
obvious question that can be raised is: can we combine these three pieces of information to
provide predictions about the flux evolution over an active region lifetime?

The answer to the question is affirmative. The importance of the peak 304 \AA\ intensity is that
it is directly linked to maximum flux in the region (Figure~\ref{fig:flux-lum}), which coincides
with the stoppage of flux emergence. From that moment, the evolution of the flux is fundamentally
governed by the diffusive motions of convection, along with the differential rotation and meridional
flows \citep[e.g. see review][]{jiang2014}. Transport by diffusion has been known for many decades
\citep{leighton1964}. In the case of a bipole, it has been shown \citep{mosher1977} that the flux
decay in one polarity can be expressed as:
\begin{eqnarray}
\phi_+ = \phi_o \erf\left({a \over \sqrt{4Dt}}\right)
\label{eqn:erf}
\end{eqnarray}
where $\phi_o$ is the initial flux, $2a$ is the separation between polarities and $D$ is the
diffusion coefficient. The expression illustrates that provided we know the diffusion that affects
the field, we only need to know the strength and the separation of the polarities at the start of
the process, to predict the decay over time. Simple as this is, however, it is insufficient to fit
the slope of the light curves in Figure~\ref{fig:ARcurves}. The limiting factor is that it
describes ideal conditions and it does not take into consideration the flux lost under the
sensitivity threshold of the instrument or the effect of the convective motions concentrating the
magnetic field into the magnetic network. This can be addressed numerically in a model like the
AFT model.

\begin{figure*}[htbp!]
\centering
\includegraphics[width=2\columnwidth]{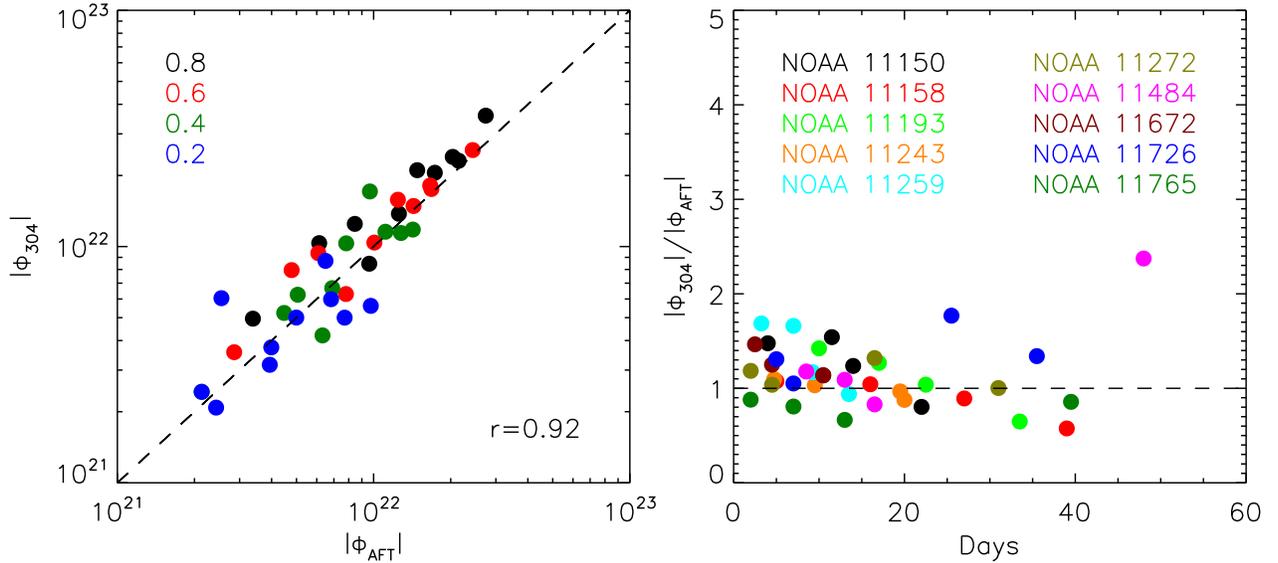}
\caption{Left: Comparison of total unsigned flux for the 304 \AA\ and AFT curves for times when $|\phi_{304}|$
reaches a fraction of the peak flux. Fractions are denoted by the colors. $r$ is the Pearson's correlation coefficient.
Right: Ratio of both fluxes as a function of time for the same instances. Those times and fluxes are highlighted with
circles in Figure~\ref{fig:SFTforecast}. Dashed lines indicate expected values for the same flux.}
\label{fig:metrics}
\end{figure*}

In Figure~\ref{fig:SFTforecast}, we show comparisons between $|\phi_{304}|$ and $|\phi_{SFT}|$
with the restriction that the assimilation of new magnetic field data is stopped when the 304
\AA\ light curve reaches its peak intensity. From that moment onward, the total flux curve is from
a field distribution created purely by the modeled transport. The plots show that the AFT model
does very well in reproducing the amount of flux decayed over one solar rotation.

In most cases the active region evolution is well matched for the entire period. In cases like 11158,
observations and model diverge as the region leaves the Earth view for a second rotation. This may be
caused by interactions with nearby flux that has not been assimilated into the model. In fact, two small
regions emerged to the East on March 19 and 26, interacted and disappeared, coincidental with drops in
the observed total flux. In cases like 11259 and 11672, a rapid decay is observed
immediately after the assimilation process is halted. This is likely caused by stochastic differences
between the actual and simulated convection patterns. As the active region flux adjusts to the new
network pattern, leading and following polarity flux are more prone to interact and cancel with one
another. After this brief adjustent  period, the decay process stabilizes and mirrors the evolution of
the 304 \AA\ flux. We note that this effect may be more pronounced for active regions with a smaller
separation distance between the leading and following polarity. For example, while 11259 and 11765 had
similar peak intensities, 11765 had about twice separation distance and did not experience a substantial
adjustment period. It should also be noted that most of these discrepancies are mitigated by continued
assimilation of magnetograms when they are avaialable.

The bottom three panels in Figure~\ref{fig:SFTforecast} (NOAA 11272, 11484 and 11726) are a special
subset in the predictions
because the peak in 304 \AA\ is reached while on the back side of the Sun and therefore we do not
have information about the flux distribution when the decay starts. This makes them excellent
cases to test how 304 \AA\ images and light curves serve as the single source of information for a
new emergence and its future decay. For these cases, we manually insert a bipolar active region in
the AFT maps when the 304 \AA\ has reached its maximum value. In the case of 11484 and 11726,
the assimilation is stopped before the actual emergence of the region.

The bipolar strength is chosen from the total unsigned magnetic flux inferred from 304 \AA, split
between the two polarities. The separation is derived from the weighted standard deviation from
the centroid of the 304 \AA\ intensity distribution over a threshold of 170 DN/s.  That distance
is shown as a blue circle in Figure~\ref{fig:separation}. We determined the relationship between
this standard deviation and the distance between polarities for 19 active regions near central
meridian and used it to estimate the polarity separation for the three regions with emergences on
the back side of the Sun where we only have information about 304 \AA. While it is possible in
some cases to also estimate the tilt angle for the two polarities based on a 2D Gaussian fit of
the 304 \AA\ intensity distribution (green elipses in the figure), the method proved unreliable in
general without further constraints.
The tilt angle ($\gamma$) was, therefore, simply estimated from Joy's law for a given latitude
$L$, $\gamma(L)=0.51|L|-0.8\degr$.  This can be a crude approximation because it is well known
that there is a wide scatter around the latitude average \citep{Wang_etal89}.

How reliable are these predictions? The flux can be consistently predicted within a factor of 2.
Figure~\ref{fig:metrics} shows a comparison
of 304 \AA\ and SFT fluxes for times when the curves reach a fraction (0.8, 0.6, 0.4 and 0.2) of the peak
in the $|\phi_{304}|$ smoothed light curve. When plotted as a function of time since peak intensity,
precision remains rather constant, with only a few points suggesting that it may slowly
worsen with time. This does not necessarily imply that the modeling of the flux decay is inaccurate
at longer times. Given that many points lie well within that factor of 2 range, it can just be a reflection
of the increased chance of other contributions not considered in the model, such as new emergences,
affecting the light curves.

These results confirm that, despite the
approximations, 304 \AA\ measurements can be sufficient when coupled with a SFT model, to establish
how the total unsigned flux of an active region changes over long time scales. Furthermore, a SFT
model that combines assimilation of nearside data with far side data from 304 \AA\ measurements
should be able to provide a more complete picture of the magnetic field configuration of the
entire Sun. This has significant implications for space weather predictions, such as solar irradiance,
solar wind, and coronal field models.

\section{Conclusions}

We present a study of the long term evolution of active regions taking advantage of the full Sun
perspective given by the combination of EUVI/STEREO and AIA/SDO images. We find that the 304
\AA\ light curves for isolated active regions of different sizes are scalable by their peak
intensities in 304 \AA. We show that these 304 \AA\ intensities are directly linked to the total
magnetic flux in the regions and use the AFT model to investigate the flux decay over several
rotations. We find that the AFT model provides a realistic prediction, within a factor of 2,
for the total flux evolution when it is coupled with information about the time of maximum
flux in the region. Finally, we show that when magnetic field data is not available, 304 \AA\ images
can provide sufficient information about the current stage of active region development to make those
predictions plausible. These results can have implications in the exploitation of the data from
current and future missions observing the far side of the Sun, and their potential use for predictive
science in combination with state-of-the-art magnetic flux transport models. The applicability
of this technique needs to be demonstrated in active regions inside large active region
complexes, which are frequently observed around the peak of solar activity.


\acknowledgments We would like to thank the referee for insightful comments that helped
improve the paper. IUU acknowledges funding from the NASA grant NNX13AE06G.
HPW's participation was supported by CNR. The SECCHI data are produced by an international
consortium of the NRL, LMSAL and NASA GSFC (USA), RAL and U. Bham (UK), MPS (Germany),
CSL(Belgium), IOTA and IAS (France). AIA and HMI data are courtesy of NASA/SDO and the AIA and HMI
science teams. IUU and HPW would like to thank Neil Sheeley for many helpful conversations.
IUU also acknowledges useful comments from William T. Thompson about the EUVI-AIA corrections
available in the SolarSoft STEREO beacon directories.

\bibliography{ms.bib}

\end{document}